\documentclass[useAMS,usenatbib]{mn2e}
\usepackage{graphicx}

\def\la{\raise.5ex\hbox{$<$}\kern-.8em\lower 1mm\hbox{$\sim$}}
\def\ma{\raise.5ex\hbox{$>$}\kern-.8em\lower 1mm\hbox{$\sim$}}

\def\msol{M$_{\odot}$ }

\def\kms{$\rm km\, s^{-1}$}
\def\cm3{$\rm cm^{-3}$}
\def\Ts{$\rm T_{*}$~}
\def\Vs{$\rm V_{s}$~}
\def\n0{$\rm n_{0}$}
\def\B0{$\rm B_{0}$}

\def\erg{$\rm erg\, cm^{-2}\, s^{-1}$}
\def\mum{$\mu$m~}

\def\L12{L$_{12\mu m}$~}
\def\F12{F$_{12\mu m}$~}

\def\Hb{H${\beta}$}
\def\Ha{H${\alpha}$}
\def\Haa{H${\alpha}_{calc}$~}
\def\Ly{Ly$\alpha$~}
\def\La{L$_{H\alpha}$~}

\def\cf{{\it cf}}

\title[Intermediate redshift galaxies]{Modelling the spectra of galaxies at
redshift 0.4$<$z$<$1.15:
the gas physical conditions and  the element abundances}
\author[M. Contini]{M. Contini  \\
School of Physics and Astronomy, Tel Aviv University, Tel Aviv 69978, Israel \\
}

\begin{document}

\date{Accepted: Received ; in original form 2010 month day}

\pagerange{\pageref{firstpage}--\pageref{lastpage}} \pubyear{2009}

\maketitle

\label{firstpage}

\begin{abstract}
We revisit the spectra observed  by Ramos Almeida et al. (2013) 
from an homogeneous sample of galaxies  at 0.27$<$z$<$1.28
with the aim of
finding out the characteristics of the single objects by consistent modelling. 
In particular we investigate the trend of  
the most significant physical  parameters  and of the element abundances with  redshift.
The observed corrected line ratios cover a relatively large range. 
Nevertheless  the calculated  physical conditions
are similar to those of  active galactic nuclei and of starburst  galaxies.
For some of the galaxies the dominant  photoionizing source is ambiguous.
The  N/H and O/H relative abundances are close to the solar ones but show a dip 
at z$\leq$0.7.

\end{abstract}


\section{Introduction}

 The mutual interaction between the nuclear activity and  starbursts in galaxies 
is generally investigated in order to understand star formation processes and epochs
as well as the nuclear characteristics.
The formation of stars within galaxies at  various redshifts  depends on  
the interstellar medium (Spaans \& Carollo 1997).
They claim that star formation  is connected  with the heating of gas and dust by the ionizing 
radiation flux due to feedback mechanisms for instance  from supernovae
and stellar winds. However collisional processes cannot be neglected. 

During major mergers of spiral galaxies at higher redshifts
(Springel  et al 2005)
the  collision and mixing of galaxies debris trigger nuclear gas inflow, which leads to
  energetic starbursts and  black hole accretion.
 The  mutual feedback of starbursts (SB)
and  active galactic nuclei (AGN)   modify the galaxy \Ha~ luminosity
 because both the SB and the   AGN  contribute  to the  total observed \La.
The  \Ha~ emission line flux  and the  flux  of  lines corresponding to  the other elements
emitted  from  the gas  close to the SB and
 from  the gas  ionized by the  AGN radiation flux  are summed up
in the  characteristic  spectra observed at Earth (Contini 2013a).

In this paper we would like to investigate  the dominant properties
of galaxies at intermediate redshift through the analysis of their spectra,
both line and continuum.
The higher the  redshift the harder is to observe enough spectral lines 
as to allow a reliable  interpretation of the spectra.

Recently, Ramos Almeida et al (2013, hereafter RA13) presented the line and continuum spectra
of a sample of galaxies at 0.27$<$z$<$1.28. 
The observations which  refer to different galaxy types
all containing an AGN cover a  
relatively extended intermediate redshift range, 
providing a rare occasion to investigate the conditions of the gas in AGNs and close to SBs
with  a high enough precision, as it is generally done for local galaxies.

The analysis of the data  presented by  RA13  yields interesting results
about  star formation rates (SFR). 
In this paper, however,  we would like to  focus on the 
gas properties in single objects as a function of the redshift.

Whether the AGN or the starburst SB dominates in  each galaxy
is  investigated in this paper by modelling the observed spectra by two different kinds of models : 
those  characterised by a power-law  photoionising  flux 
 and those  characterised by a black-body radiation flux.
It was suggested that galaxies at intermediate and high redshift  are mergers
originating from collision of galaxies at earlier epochs.
Therefore, both  photoionization and  shocks are accounted for in the calculation
of the spectra.

Recently, the spectral energy distribution (SED) of the continuum has been enriched  by 
far infrared observations  which
 contribute to the understanding of  the galaxy conditions relative to  dust.
The shock effect can be recognized from  the maximum frequency and intensity of the reprocessed dust radiation peak
in the infrared and of the bremsstrahlung   at
 high frequencies of the continuum SED.
 The models are selected by  cross-checking the 
fit of  the continuum SED data  by the fit of the observed  line ratios.
However, only a few objects at high z  show enough data  throughout the SED
 to constrain the models, as was demonstrated for local
galaxies.

We start by  modelling   line and continuum  spectra observed from  galaxies selected from the RA13 sample 
 at 0.4$<$z$<$1.15. 
They were chosen among those showing the largest number of significant lines.
Moreover, RA13 presented also near-IR spectroscopic observations which  were used to 
classify the observed SEDs.

In Fig. 1 the  line ratios observed from the selected galaxies are  plotted  throughout a 
Baldwin-Phillips-Terlevich (BPT) diagram (Baldwin et al. 1981). 
Fig. 1 shows that  the   line ratio  ensemble 
crosses the lines separating AGNs from the HII regions. The grids of calculated line ratios for local merger galaxies
 by Contini (2012a, fig. 1) show that the  domains of AGN and SB in a BPT diagram may overlap.

\begin{figure}
\includegraphics[width=8.6cm]{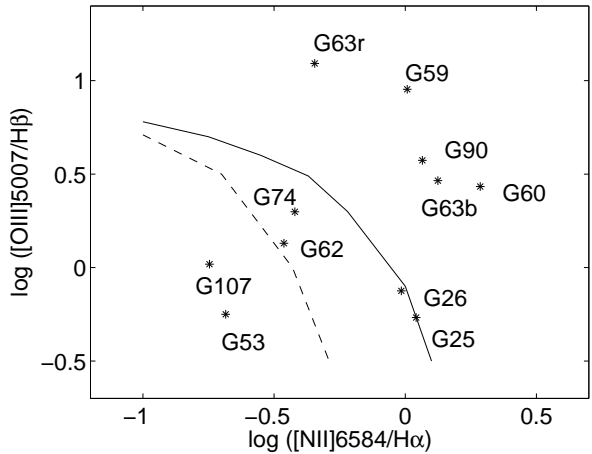}
\caption{The distribution of the  galaxies throughout a BPT diagram.
 RA13 observed line ratios corrected for extinction: black asterisks; 
the black dashed and solid lines represent the empirical separation between AGN and HII regions by Kewley et al (2001)
and Kauffmann et al (2003), respectively}
\end{figure} 

The paper is organised as  follows.
A brief description of the calculation code is given in Sect. 2.
The intermediate redshift (0.4$\leq$z$\leq$1.2) galaxies are analysed  
considering   line ratios to \Hb~  and 
continuum spectra in Sect. 3.
Discussion and concluding remarks appear in Sect. 4.

\begin{figure}
\includegraphics[width=8.6cm]{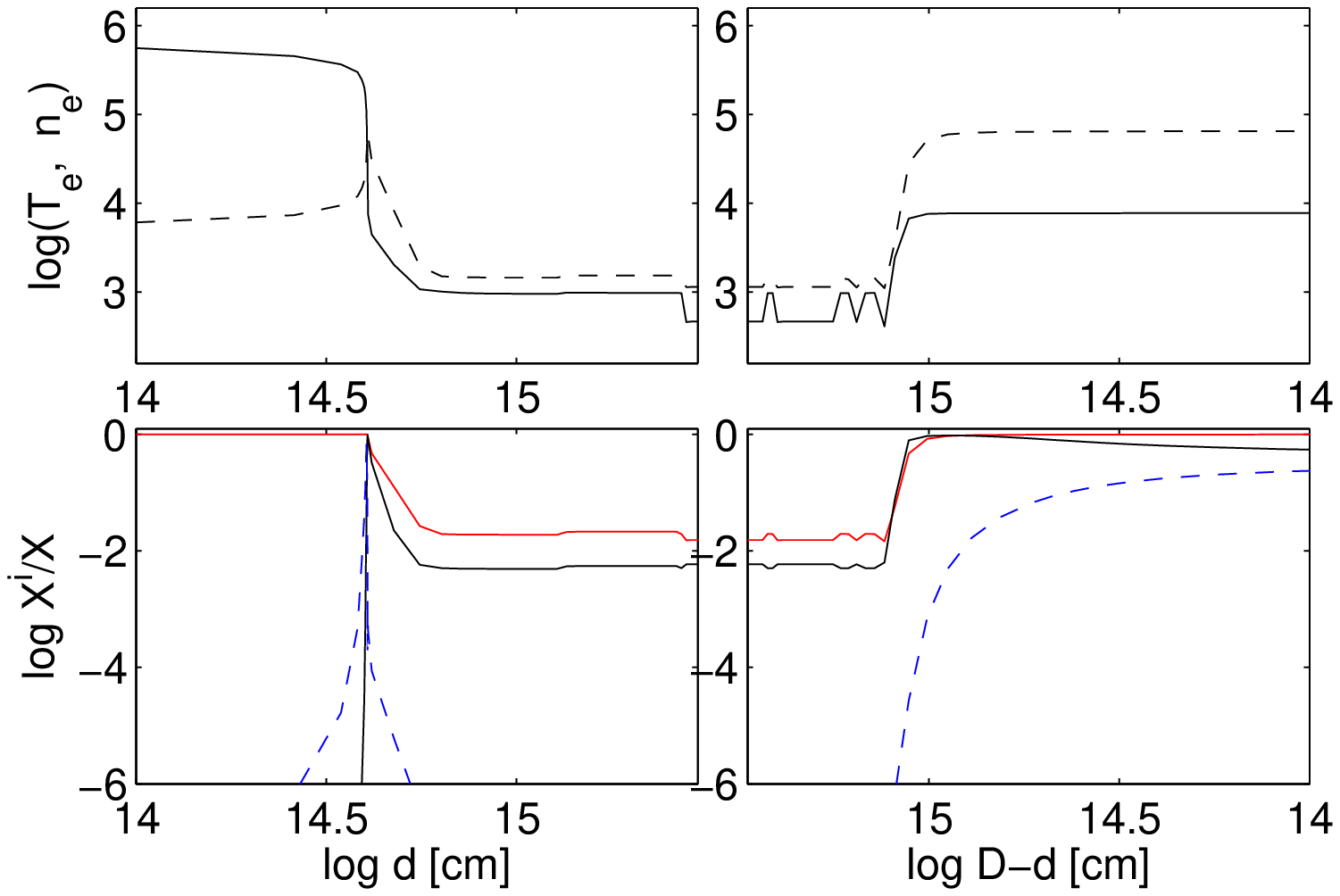}
\caption{Top: the electron temperature (black solid line) and the electron density (black dashed line)
throughout a cloud corresponding to G25 for the SB model Mz25. The cloud is ejected outwards.
Bottom: the same as in the top panel for the profile of the O$^{2+}$/O (blue dashed line),
N$^+$/N (black solid line) and H$^+$/H (red solid line)}
\includegraphics[width=8.6cm]{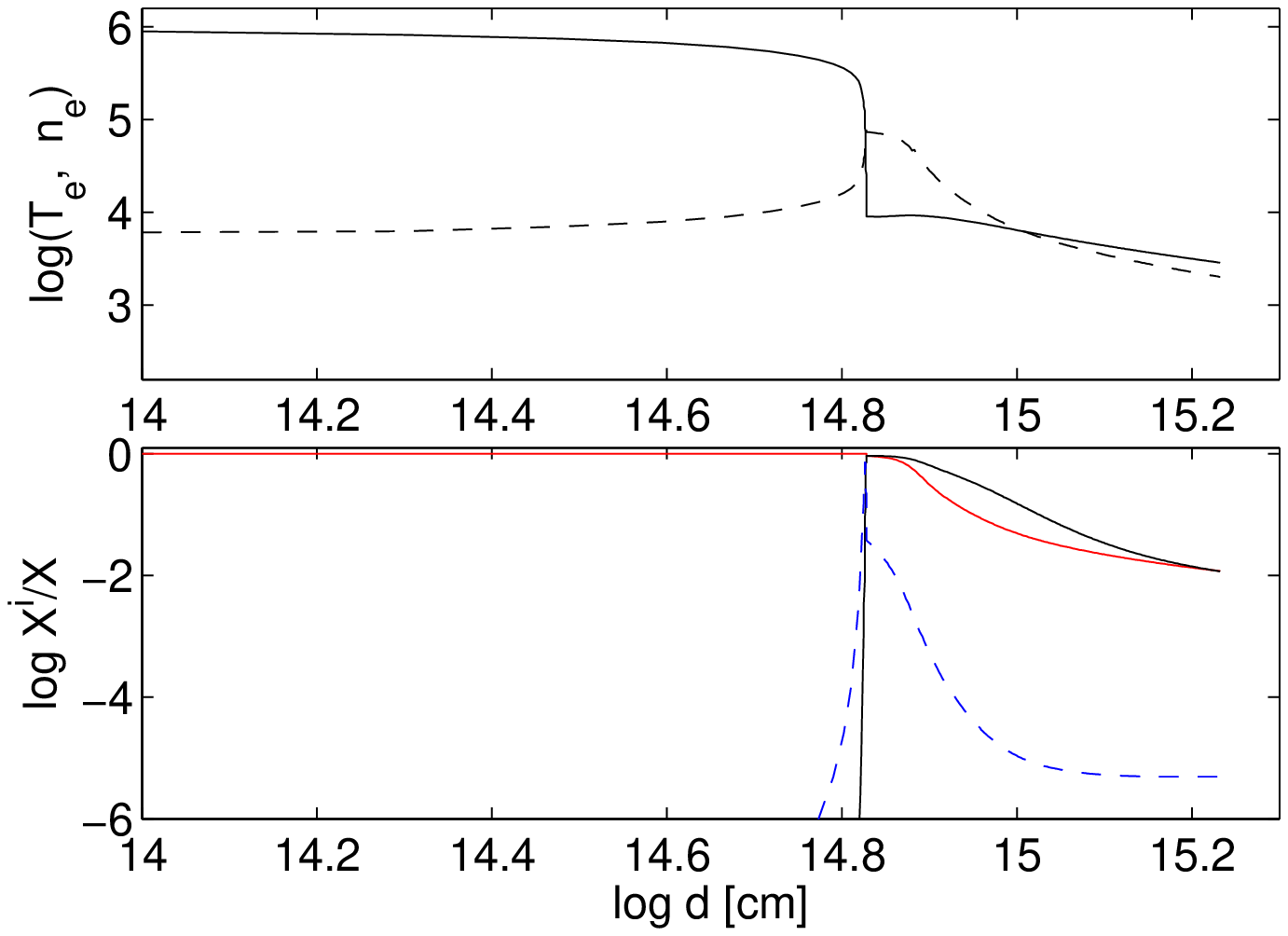}
\caption{The same as in Fig. 2 for the AGN model M25. The cloud is infalling towards the AGN}
\end{figure}

\section{The calculations}

The observed spectra   provide the tools for calculating the
physical and chemical conditions in galaxies  at each redshift.
Metallicity (in term of O/H and  N/H relative abundances) is calculated from the
[OIII]/\Hb,  [OII]/\Hb~  and [NII]/\Hb~ line ratios.
At low redshifts (z$\leq$0.1)  the calculated  relative abundances   are  generally constrained by
   many lines from different elements in different ionization levels  throughout  UV-optical-infrared spectra.
Photoionization dominated models are used to calculate the temperature of the emitting gas (e.g. Izotov et al. 2006,
Stasinska \& Izotov 2003).
We use composite models  accounting for the combined affect of photoionization from the AGN or from the SB and shocks
(e.g. Contini 2013a and references therein) which are more adapted to   deal with the collisional processes throughout
the   mergers.

The  modelling of the spectra at higher z
becomes ambiguous    when one or more key lines from both  high and low ionization levels are leaking.
Therefore, the choice of the models will be dictated by  considerations  regarding
not only   the SB or the AGN radiation spectrum, but also the  hydrodynamical regime  and the galactic cloud structure.
Moreover, the observed line ratios are often uncertain due to selection effects, calibration biases and  different
corrections from obscuration (Gunawardhana et al. 2013).
In an accompanying paper (Contini 2013b) we discuss
the reliability of the results obtained by modelling spectra showing only a few oxygen to \Hb~  line ratios.

We have adopted for the calculations of the spectra the code
  {\sc suma}\footnote{http://wise-obs.tau.ac.il/$\sim$marcel/suma/index.htm}, which
 simulates the physical conditions in an emitting gaseous cloud under the coupled effect of
photoionization from an external radiation source and shocks. The line and continuum emission
from the gas are calculated consistently with dust-reprocessed radiation in a plane-parallel geometry
(see Contini et al 2012 and references therein for a detailed description of the code).

The input parameters which characterise the model are : the  shock velocity \Vs, the atomic preshock density \n0,
the preshock magnetic field \B0. They define the hydrodynamical field and  are  used in the calculations
of the Rankine-Hugoniot equations  at the shock front and downstream. These equations  are combined into the
compression equation which is resolved  throughout each slab of the gas, in order to obtain the density
profile downstream and consequently, the temperature of the gas.
We adopt  for all the models \B0=10$^{-4}$ gauss. In fact, a higher preshock density is compensated by a
higher magnetic field which reduces compression and viceversa. We do not have enough data
  to avoid degeneration  of models   choosing  between  \n0  and \B0 effects on the density (see Contini 2009).

The input parameter  that represents the radiation field  in AGNs is the power-law
flux  from the active center $F$  in number of photons cm$^{-2}$ s$^{-1}$ eV$^{-1}$ at the Lyman limit,
if the photoionization source is an active nucleus. The spectral indices are $\alpha_{UV}$=-1.5
and $\alpha_X$=-0.7.
 $F$  is combined with the ionization parameter $U$ by
$U$= ($F$/(n c ($\alpha$ -1)) (($E_H)^{-\alpha +1}$ - ($E_C)^{-\alpha +1}$)
(Contini \& Aldrovandi, 1983), where
$E_H$ is H ionization potential  and $E_C$ is the high energy cutoff,
n the density, $\alpha$ the spectral index, and c the speed of light.

If the radiation flux  is  a black body radiation from the stars,
the input parameters are the colour  temperature of the  star \Ts
and  the ionization parameter $U$
(in  number of photons per  number  of electrons at the nebula).

The secondary diffuse radiation emitted from the slabs of gas heated by the shocks is also calculated.
The  flux  from the AGN,  from the stars and the secondary radiation flux are  calculated by
radiation transfer throughout the slabs downstream.

The geometrical thickness of the emitting nebula $D$,
the dust-to-gas ratio $d/g$, and the  abundances of He, C, N, O, Ne, Mg, Si, S, A, and Fe relative to H
are also accounted for.
 The distribution of the grain radius  downstream
is determined by sputtering,  beginning with an initial  radius of  $\sim$ 0.5-2.5 \mum.

In our models the flux from an external source can reach the  shock front (the inflow case,
indicated by the parameter $str$=0) or the edge of the
cloud opposite to the shock front  when the cloud propagates outwards from the AC or from the starburst
(outflow, indicated by $str$=1).

The fractional abundances of the ions are calculated resolving the ionization equations
considering ionization by the primary flux, the secondary flux and collisional ionization
in each slab downstream.
The  line intensity calculated in each slab  depends  strongly
on  the density,  temperature, and radiation from both the sides of the cloud.
The line intensities are integrated throughout the downstream region up to a distance $D$  from
the shock front at which the gas has reached a relatively low temperature (T$<$ 10$^3$ K) in the case of
inflow (radiation bound case) or to a distance $D$ such that all the calculated line ratios reproduce
the observed ones (matter bound case).
In the outflow case some iterations are necessary to obtain converging results.

The models  which  reproduce the  observed strong lines within 20\% and the weak lines by 50 \%
are selected from a large grid of models. The  sets of  input parameters
are  regarded as {\it results}.  The approximation of calculated to observed line ratios
 should  account for the observational
errors and for the uncertainty of the physical coefficients adopted by the calculation code.

\section{The  galaxy sample} 

RA13 presented near-IR (NIR) spectroscopic observations of 28 X-ray and mid-infrared selected sources at median redshift 
z$\sim$ 0.8 in the Extended Groth Strip :
 13 AGN dominated and 15  host-galaxy dominated. New NIR spectra referring to the continuum SEDs
of  objects at  z between $\sim$ 0.4 and $\sim$1.25 were  reported. Moreover,
they  show the spectra of an AGN subset at 0.27$<$z$<$  1.28
 including   \Ha~ and other key optical lines for each of them.

Following our usual procedure in the analysis of galaxies, we first consider  the line ratios because,
as  demonstrated by Contini (2013 and references therein), 
the continuum SEDs  less constrain  the models.

The data   reported  by RA13
in their table 3 and completed  by the data presented in their table 4  were observed by  the Long- Slit
Intermediate Resolution Infrared Spectrograph (LIRIS) at the 4.2m William Herschel Telescope (WHT) from the
Deep Extragalactic Evolutionary Probe 2 (DEEP2).
We report in Table 1 the flux of the lines  and the  FWHM (indicated as F/W0 for each galaxy) of the selected objects. 
The observed fluxes are in 10$^{-16}$ erg s$^{-1}$ cm$^{-2}$
and the FWHM in \kms. We do not  show the errors given by DIPSO, for sake of clarity. Brackets indicate that 
measurements are from DEEP2 optical spectra.  Typical line width uncertainty is 25-30 \%.

We have  selected   9 out of the 28 galaxies of the RA13 sample which
 show  enough line ratios for a  suitable modelling  without any  risk of degeneracy (see Contini 2013b).
G62 was added to the sample because  the characteristic parameters are different from those of  the other objects, 
being a simple starburst.

 We have corrected the line flux ratios  to \Hb~ from extinction adopting the formula reported by
Osterbrock (1974, eq. 7.6)  and  \Ha$_0$/\Hb$_0$ (at the  nebula)  = 3 :

\noindent
 \Ha/\Hb=(\Ha$_0$/\Hb$_0$) 10$^{-c (f(H\alpha)-f(H\beta))}$, 

\noindent
where 
f($H\alpha)-f(H\beta)$  is the standard interstellar extinction curve  
(Osterbrock 1974, fig. 7.1).

The spectra lacking the \Ha ~ data (e.g. the red component of the  galaxies G53 and  G74)   were not  modelled because 
the  line flux ratios could not be corrected. Both blue and right  component spectra of G63 are  accounted for.

\begin{table*}
\caption{observed line fluxes from Ramos Almeida et al. (2013)}
\tiny{
\begin{tabular}{ccccccccccccccccccc} \hline  \hline
%\ ID& z  &\Ha/W   &[NII]6584/W     &\Hb/W   &[OIII]4959/W      &[OIII]5007/W    &[SII]6732/W   &[OII]3727/W   &[OI]6300/W&[OII]7320/W  \\
\ ID& z  &\Ha   &[NII]6584     &\Hb   &[OIII]4959      &[OIII]5007    &[SII]6732   &[OII]3727   &[OI]6300&[OII]7320  \\
\   &    &F/W   &F/W     &F/W   &F/W      &F/W    &F/W   & F/W  &F/W &F/W \\ \hline
\ G25&0.761&1.48 /$\leq$340&1.16      / $\leq$340&   [0.25] / [140]     &[:0.07  / [130]      &[0.17]  / [130] & ... / ... & ,,, / ... & ... / ... \\
\ G26&0.808&1.03 /$\leq$280&0.72      / $\leq$250& [0.09] / [70]        &[:0.05] / [70]       &[0.10] / [70]   & ... /... & ... / ... & ... / ... \\
\ G53&0.72 &1.47 / $\leq$340&0.22     / $\leq$340 & [0.08] / [$\leq$75]&[:0.06] / [$\leq$70] & [:0.06]/ [$\leq$70] & 0.31    /$\leq$340 &.... /.... &.... /....&.... /....\\
\ G59&0.465&1.42 / 230      &1.05     / 230        & [0.32] / [220]     &[1.09]  /[210]      &[3.46]  / [210]      &....     /....  &..../.... &..../....&1.28/ 210\\
\ G60&0.484&1.14 / 230      &1.6      / 230        & [0.34] / [170]     &[0.30]  /[160]      &[1.02]  / 160]      &....     /....      &....     / ....   &.... /....&..../....\\
\ G62&0.902&2.87 /50       & 1.1      /50         &  :0.97  /40        & :0.97   /40        & 1.67    /40        &....      /....     &[0.50]   / 160   &.... / ....&...../....\\
\ G63b&0.482&4.82 / 590      &4.67     / 590        & [0.07] / [880]     & [0.23 / [850]      &0.69    / [840]     &....    /....       &....     / ....  & 2.09 / 470&..../....\\
\ G63r&0.482&2.39 / 230      &1.33     / 230        & [0.10] / [180]     & [0.52 / [310]      &1.57    / [310]     &....    /....       &....     / ....  & 2.09 / 470&..../....\\
\ G74&0.551&5.25 / 280      &1.29     / 280        & [0.53] / [120]     & [0.40] / [65]      &1.2     / [65]      &0.86    / 270       &....     / ....  & ..../ ....&..../....\\
\ G90&1.148&19.4 / 640      &16.4     / 640        & 2.58   / 780       & 4.78   / 760       &14.3    / 750       &....    / ....      &1.63     / 490   & ..../ ....&..../....\\
\   &     &                &                      &                    &                    &                    &                    &[0.51]   / [320] & ..../ ....&..../....\\
\ G107&0.671&2.6 / 90       &1.22     / 90         & [0.37] / [$\leq$90 & [:0.37] / [$\leq$85]& [:0.37]/ [$\leq$85]&....   / ....      & ....    / ....  & .... / ....&..../....\\ \hline
\end{tabular}}
 
\end{table*}

\begin{table*}
\caption{Comparison of calculated with observed line ratios to \Hb for RA13 sample}
\begin{tabular}{ccccccccccccccc} \hline  \hline
\ line   & \Ly &NV   &  CIV  & CIII]   & CII]  & MgII       & [OII]    & HeII  & [OIII] & [OI]    & [NII]  & [SII]  & [OII] &z\\
\ $\lambda$&1215&1240&  1500+& 1909+   & 2326+ &  2789      & 3727+    & 4686  & 5007+  & 6300+   & 6584+  & 6720+  & 7320+ & \\ \hline
\ G90 corr&  -   &-   &  -    &  -      & -     & -          & 1.26     & -     & 5.     & -       & 3.37   & 1.26   & -     & 1.148 \\
\ M90    & 46.7 &11. & 19.   & 8.7     & 4.6   & 1.53       & 1.3      & 0.11  & 5.2    & 0.42    & 3.3    & 0.2    & 2.3   &  \\
\ Mz90   & 28.4 &0.13& 0.03  & 0.05    & 0.07  & 0.6        & 0.9      & 0.0026& 4.5    &0.018    & 3.     & 1.3    & 0.1   &  \\
\ G62 corr    & -    & -  &  -    &   -     &  -    &   -        & 0.5-0.24 & -     & 2.65   &  -      & 1.1    & -      & -     & 0.902\\
\ Mz62   & 23.5 &0.0003&0.56 & 0.75    & 0.12  & 0.32       & 0.56     & 0.005 & 2.86   &0.0003   & 0.71   & 0.03   & 0.01  &  \\
\ G26 corr&   -  & -  &  -    &  -      &  -    &   -        &  -       &  -    & 1.     & -       & 2.8    & -      &  -    &0.808 \\
\ M26    & 35.6 &2.0 & 5.9   & 2.55    & 1.74  & 1.65       & 0.75     & 0.1   & 0.99   & 0.26    & 3.     & 0.34   & 0.48  & \\
\ Mz26   & 31.9 &0.2 & 0.25  & 0.16    & 0.24  & 0.58       & 0.47     & 0.0004& 0.94   & 0.07    & 3.     & 0.3    & 0.22  & \\
\ G25 corr&  -   &-   &  -    &  -      &  -    &   -        &  -       &  -    & 0.72   & -       & 3.19   & -      & -     & 0.761  \\
\ M25    & 37.6&2.0  & 5.67  & 2.33    & 2.    & 2.1        & 0.66     & 0.11  & 0.77   & 0.89    & 3.3    & 0.95   & 0.45  & \\
\ Mz25   & 31.9&0.23 & 0.27  & 0.17    & 0.25  & 0.56       & 0.37     & 0.0004& 0.8    & 0.063   & 3.1    & 0.3    & 0.48  & \\
\ G53 corr& -   &-    &  -    &  -      &  -    &  -         &  -       &  -    & 0.75   & -       & 0.6    & 0.04   &  -    &0.72 \\
\ M53    & 36.9&1.2  & 12.8  &  4.4    & 2.12  & 1.4        &  0.37    & 0.05  & 0.83   & 0.48    & 0.58   & 0.06   & 0.3   & \\
\ Mz53   & 32.4&0.05 & 0.39  & 0.25    & 0.46  & 1.15       & 0.48     & 0.0006& 0.8    & 0.09    & 0.58   & 0.02   & 0.21  & \\
\ G107 corr&  - & -   &  -    &  -      &  -    &  -         &  -       & -     &$<$1.39 & -       & 0.52   & -      & -     &0.67  \\
\ M107   & 33.6&0.016& 4.5   & 2.6     & 2.14  & 2.15       & 1.4      & 0.13  & 0.75   & 0.49    & 0.63   & 0.12   & 0.29  & \\
\ Mz107  & 29.5&0.0017&0.35  & 0.28    & 0.36  & 1.1        & 1.46     &0.00032& 1.16   & 0.078   & 0.5    & 0.06   & 0.28  & \\
\ G74 corr& -   & -   &  -    &  -      & -     & -          &  -       & -     & 1.8    & -       & 1.     & 0.45   &  -    & 0.551 \\
\ M74    & 37.3&1.6  & 6.65  & 3.      & 1.2   & 1.6        & 0.86     & 0.06  & 1.76   & 1.      & 1.6    & 0.4    &  1.   & \\
\ Mz74   & 32.8 &0.04& 0.096 & 0.1     & 0.23  & 0.7        & 0.4      &0.00013& 1.7    & 0.22    & 1.17   & 0.45   &  0.19 & \\
\ G60 corr&  -  &-    &  -    &  -      &  -    &  -         &  -       &  -    & 3.62   & -       & 5.6    & -      & -     &0.484  \\
\ M60    & 32.4&0.74 & 1.86  & 1.28    & 1.34  & 1.35       &  3.5     & 0.177 & 3.6    &0.58     & 5.07   & 0.6    & 0.85  & \\
\ Mz60   & 30.6&0.096& 0.11  & 0.12    & 0.127 & 0.8        & 0.49     & 0.0004& 3.1    &0.024    & 2.08   & 0.44   & 0.13  & \\
\ G63b corr & -  &-    &  -    &  -    &  -    &  -         & -        &  -    & 3.89   &0.19     & 3.86   & -      & -     & 0.482  \\
\ M63b    & 35.8&3.8  & 8.34  & 4.19    & 2.22  & 1.18       &2.        & 0.13  & 3.85   & 0.21    & 3.6    & 0.26   & 1.2   & \\
\ Mz63b   & 31. &0.012& 0.026 & 0.1     & 0.17  & 0.77       & 0.59     & 0.047 & 3.8    & 0.006   & 3.84   & 0.2    & 0.2   & \\
\ G63r corr & -  &-    &  -    &  -      &  -    &  -        & -        &  -    &16.5    &3.18     & 1.31   & -      & -     & 0.482  \\
\ M63r    & 28.0&0.27 & 1.47  & 2.1     & 1.25  & 1.04       &5.8       & 0.31  & 15.6   & 0.08    & 2.11   & 0.44   & 0.73  & \\
\ Mz63r   & 36.3&0.03 & 2.64  & 1.69    & 0.25  & 11.6       &1.        & 0.97  & 16.3   & 0.12    & 1.38   & 2.6    & 0.22  & \\
\ G59 corr&  -  &-    &  -    &  -      & -     & -          &  -       &  -    &  12.   &  -      & 2.95   & -      & 2.5   & 0.465\\
\ M59    & 51.6&12.3 &  42.  &  20.    &  7.5  & 1.1        & 6.       & 0.1   & 11.9   & 0.28    & 3.2    & 0.6    & 3.2   & \\
\ Mz59   & 34.7&0.04 &  0.7  &  0.85   & 0.29  & 6.57       & 0.93     & 0.78  & 12.6   & 0.26    & 2.5    & 1.3    & 0.33  & \\ \hline
\end{tabular}
\end{table*}

\begin{table*}
\caption{The physical parameters and relative abundances in AGN  models for the RA13 sample}
\begin{tabular}{ccccccccccc} \hline  \hline
\ model &  \Vs   &  \n0 &  $F$       & $D$    &O/H       & N/H      & \Hb abs & n$_H$  & z \\
\       &   \kms & \cm3 & units$^1$  & 10$^{16}$ cm& 10$^{-4}$&10$^{-4}$ &  \erg    & 10$^4$ \cm3   & - \\  \hline
\ M90   & 700    & 800  &2.        & 0.2   &6.6   &1.2   & 0.016   & 10.   &1.148\\
\ M26   & 200    & 1300 &2.        & 0.00584&3.    &0.8  & 0.043   & 5.5  &0.808\\
\ M25   & 250    &1300  &2.        & 0.17 &3.    &0.8  & 0.054   & 7.2  &0.761\\
\ M53   & 350    &1000  &0.8        & 0.37 &2.    &0.2  & 0.03    & 7.   &0.72\\
\ M107  & 100    &1000  &0.8        & 1.67&2.    &0.1   & 0.023   & 1.6  &0.67\\
\ M74   & 280    &1600  &2.       & 0.073 &6.6   &0.6   & 0.047   & 11. &0.551\\
\ M60   & 200    &650   &2.       &0.18  &6.6   &1.    & 0.045   & 2.   &0.484\\
\ M63b   & 600    &500   &2.       &0.3   &6.6   &1.    & 0.02    & 4.5  &0.482\\
\ M63r   & 200    &330   &2.       &0.6   &6.6   &0.5   & 0.03    & 0.74 &0.482\\
\ M59   & 260    & 600  &0.5         & 0.145&    6.6   & 0.7 & 0.0024 & 4.5  &0.465\\ \hline
\end{tabular}

$^1$ in 10$^{10}$ photons cm$^{-2}$s$^{-1}$eV$^{-1}$ at the Lyman limit
%\end{table*}

%\begin{table*}
\caption{The physical parameters and relative abundances  in SB  models for the  RA13 sample}
\begin{tabular}{ccccccccccc} \hline  \hline
\ model &  \Vs   &  \n0 &  \Ts  & log $U$   & $D$    &O/H       & N/H       & \Hb abs & n$_H$  & z \\
\       &   \kms & \cm3 & 10$^4$ K    & -  & 10$^{16}$cm  &10$^{-4}$ &10$^{-4}$  & \erg & 10$^4$ \cm3  & - \\ \hline
\ Mz90  & 400    & 300  &4.3  & -0.12     & 2.  & 8.4      & 3.2       & 0.54    & 1.4  & 1.148\\
\ Mz62  & 80     & 50   & 4.5 & -1.22     & 80.   & 2.     & 1.2          & 0.003.6 & 0.0167   & 0.902\\
\ Mz26  & 200    & 1300  & 4.  & -1.1     & 0.6  & 6.6      & 2.        & 0.63    & 5.9  & 0.808\\
\ Mz25  & 200    & 1300  & 4.  & -1.15    & 0.6  & 6.6      & 2.        & 0.58    & 5.9  & 0.761\\
\ Mz53  & 350    & 1000  & 4.  & -0.96    & 1.   & 5.       & 0.3       & 0.69    & 7.3  & 0.72\\
\ Mz107 & 100    & 1000   & 4.  & -1.52    &1.  & 6.6      & 0.2       & 0.18    & 1.86 & 0.67 \\
\ Mz74  & 280    & 1000  & 4.  & 0.48     & 100.     & 6.6      & 1.        & 1.85    & 5.8  & 0.551\\
\ Mz60  & 250    & 750   & 4.  & -0.30    & 0.9  & 6.6      & 2.        & 0.93    & 3.4  & 0.484\\
\ Mz63b  & 560    & 500   & 4.  & -0.30    & 0.6  & 8.       & 3.5       & 0.87    & 4.   & 0.482\\
\ Mz63r  & 250    & 400   & 6.  & 1.78     & 8    & 6.6      & 1.5       &1.54     & 1.3  & 0.482\\
\ Mz59  & 280    & 600  & 7.  & 1.08      & 9.  & 7.6      & 2.        & 1.92    & 2.7  & 0.465\\ \hline
\end{tabular}
\end{table*}

\subsection{The modelling procedure}

At the first glance,  RA13 table 3 shows that the FWHM are relatively high ($\geq$ 200 \kms), except for G62.
The velocities are in the range of those observed in the 
NLR of Seyfert 2 galaxies (200-1000 \kms Contini \& Viegas 2001) and of LINERs (100-300 \kms Contini 1997).
Most probably the galaxies at relatively high z are the result of mergers, 
justifying the use of composite models which account  for  the radiation flux and for the   shocks. 

First, we have modelled the spectra of the  galaxies showing  the largest number of lines, 
checking  whether they  were enough to constrain the models (Contini 2013b). 
Then, we have  reproduced the spectra observed  from the other objects where at least \Ha, \Hb, [OIII] 5007+  
(the + indicates that [OIII] $\lambda$5007 and $\lambda$4959 are summed up) and 
[NII] 6548+  (the + indicates that [NII] $\lambda$6548 and $\lambda$6584 are summed up) were observed.

The calculations show that \Ha/\Hb $\sim$ 2.8- 3 refers to a gas in normal conditions of density ($\leq$ 10$^6$ \cm3)
 and temperature ($\leq$ 10$^3$ - 10$^6$ K) (Osterbrock 1974).

Constraining the  shock velocity by the FWHM we chosed the  other  input parameters  such   as to obtain the best fit to
the observed [OIII]/\Hb. The other line ratios (e.g. [NII]/\Hb) are consistently calculated by the code adopting as a first 
guess the solar abundances (Allen 1976). We then changed the N/H relative abundance in order to fit
 both  [OIII]/\Hb~ and [NII]/\Hb.  Recall that the line intensity fluxes affect the cooling rate downstream
and in turn also the [OIII]/\Hb~ result. So  [OIII]/\Hb~ and [NII]/\Hb~  were cross checked changing the input parameters
until the best  fit was achieved for  both  the  line ratios.
Some  spectra in the sample report the data for
other lines, such as [OII] 7320+, [SII]6717+6731, or [OII] 3727+ which were used
to further constrain the models.

 To understand the trend of the line ratios emitted from the clouds, we show in
Figs 2 and 3  the profile of the electron density, of the electron temperature and of the O$^{++}$/O, N$^+$/N
and H$^+$/H ion fractional abundances across the clouds  in the galaxy G25.
In Fig. 2 the cloud is divided in two  halves with logarithmic X-axis scales (direct in the left panel and 
reverse in the right panel) in order to show the critical regions at the edges of a cloud
propagating outwards from the starburst. The shock front is on the left of the left
panel, while the black-body flux from the star reaches the right edge of the right panel.
Fig 3. shows the case  of a cloud infalling towards the AGN. The shock front is on the left and the
power-law radiation flux from the AGN reaches the very shock front.
 Figs. 2 and 3 show that the [OIII] lines are  result from integration throughout regions of gas 
at different temperatures and densities.

The results will change  by  varying even slightly one of the input parameters. 
 Spectra containing only   [OIII]5007+/\Hb~ and [OII]3727+/\Hb~
 (the + indicates that the doublet is summed up)
could lead to degeneracy of results, in particular  for  the O/H  relative abundance (Contini 2013b).

In Table 2 we compare  the  calculated with observed line intensity ratios to \Hb, corrected for extinction.
We have ordered the galaxies by their redshift.
 Below each  row containing the observed line ratios two  rows showing the model results follow.
The  row next to the data contains  models M25-M107 referring to the AGN (Table 3). In the next row
models Mz25- Mz107 refer to the SB (Table 4).

The input parameters which lead to the best fit of the line ratios are presented in Tables 3 and 4 for 
the AGN and for the SB models, respectively.
Tables 3 and 4 show that the relative abundances of O/H and N/H  vary from object to object. 
The abundances relative to
H for the other elements whose lines do not appear in the spectra are assumed to be solar (Allen 1976).
In columns 8 and 9 of Tables 3 and 4 we report for each galaxy the calculated  \Hb~ absolute flux  
 and the hydrogen density which results after compression of the gas downstream.

We have included in Table 2  some of the most significant line ratios in the optical and UV ranges,
even if they are not observed. They can give some  information of 
the \Ly line intensities  in galaxies at  relatively high z.
Interestingly, the range of the \Ly/\Hb~ line ratios calculated by the models fitting the RA13 sample at z$\sim$ 0.8
is similar to that of the observations in the different location of Mrk 3 (z=0.0135) by Collins et al (2009).
Notice that the OV 1215 line can be strongly blended with \Ly.
The ratio of the CIV/CIII]/CII] lines are  significant as well.
Our results show that C/H  is eventually depleted from the gaseous phase, because the calculated CIV/\Hb
seems extremely high adopting a solar C/H.

\begin{figure*}
\includegraphics[width=8.6cm]{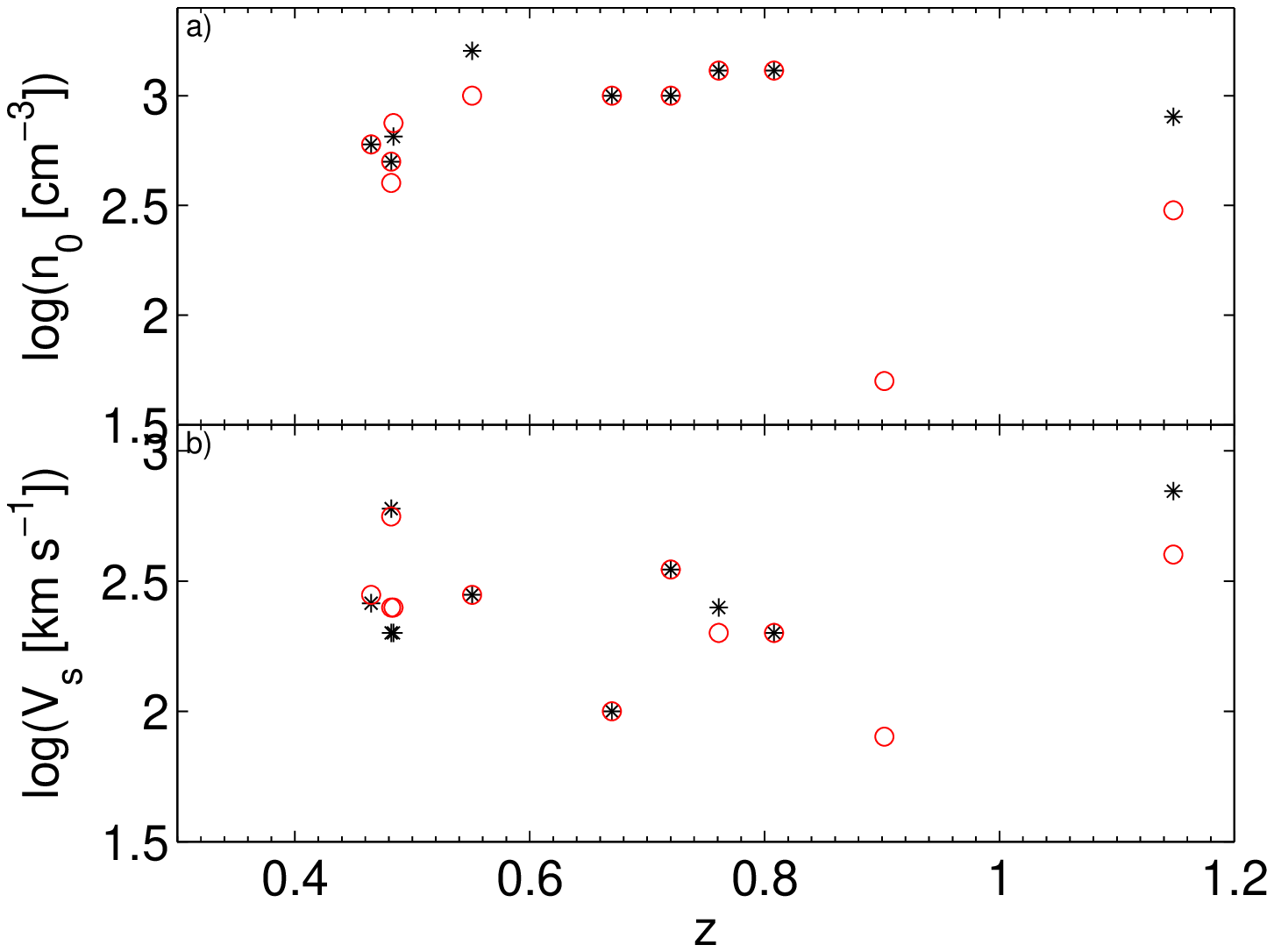}
\includegraphics[width=8.6cm]{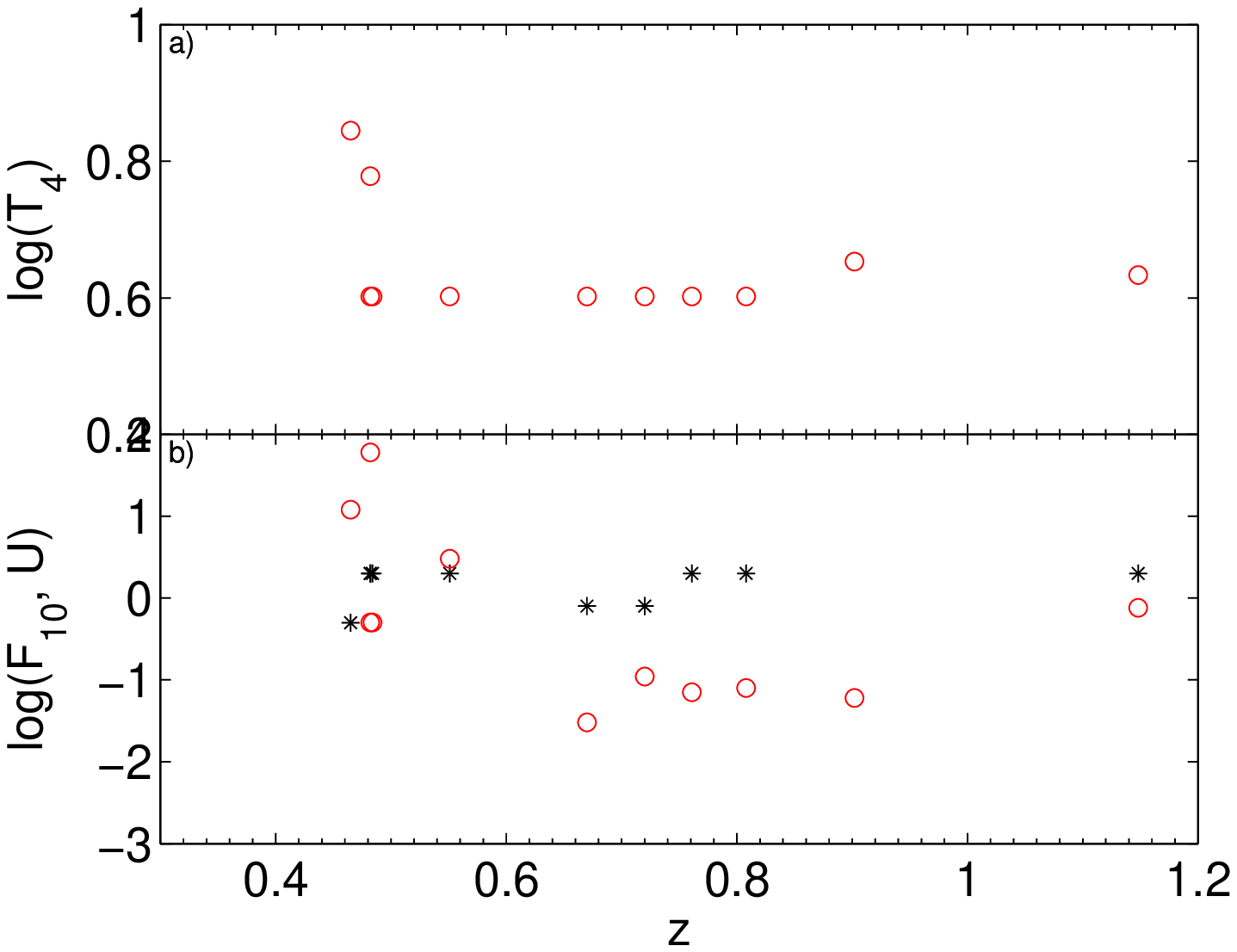}
\includegraphics[width=8.6cm]{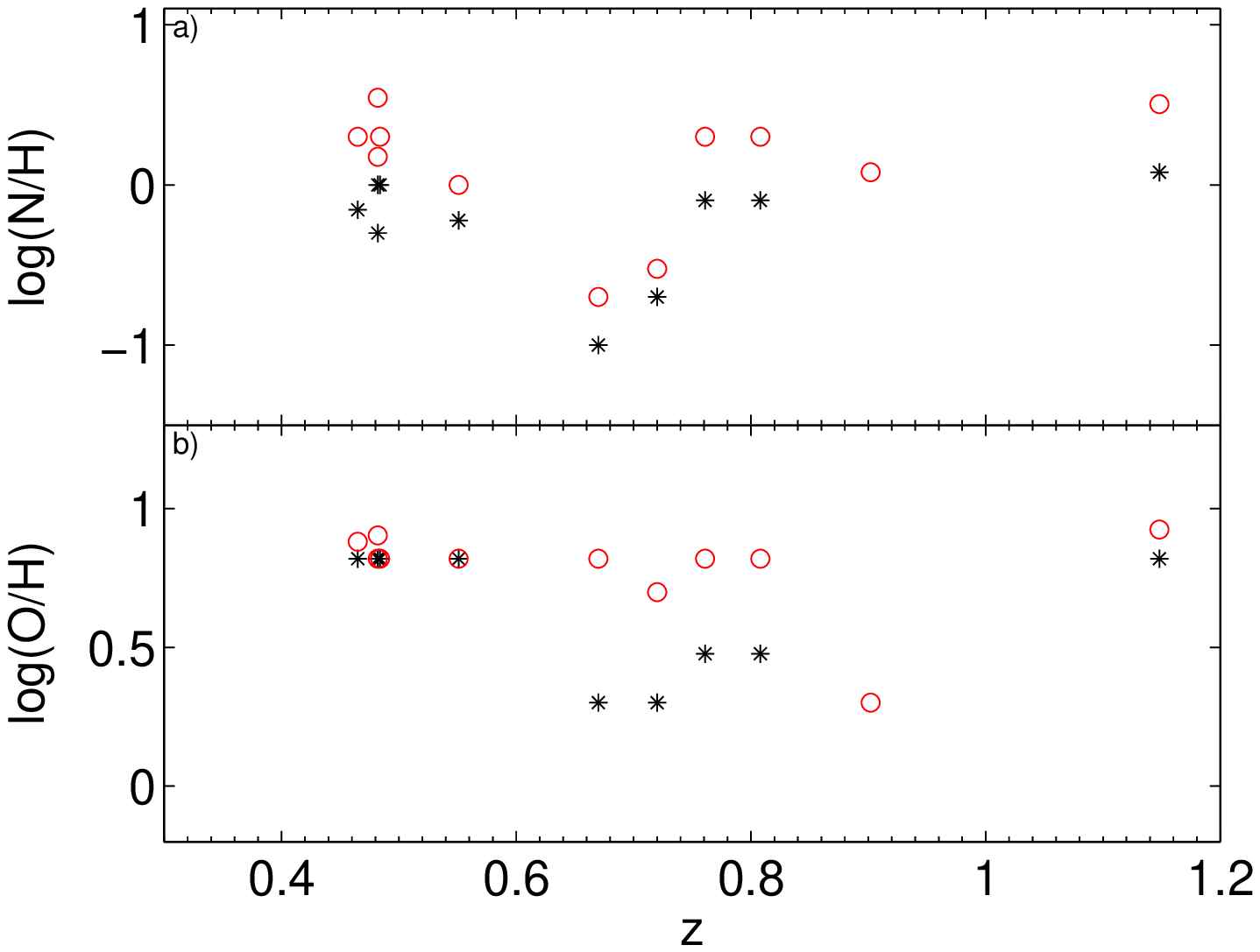}
\includegraphics[width=8.6cm]{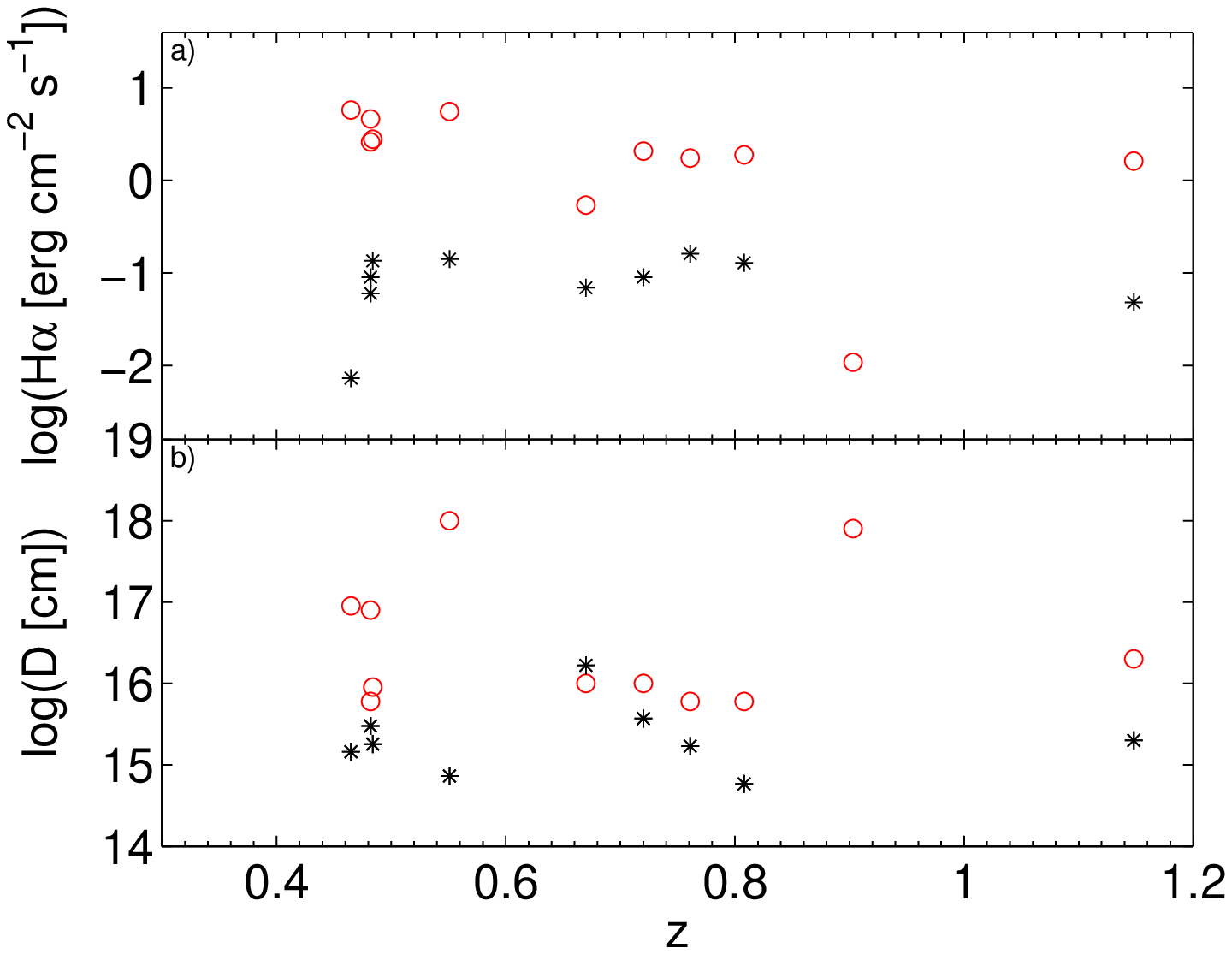}
\caption{The parameters selected by modelling the RA13 galaxy sample.
Red circles: models calculated for starbursts;
black asterisks: models calculated for  the  AGN.
Top left : \n0 and \Vs; top right : the radiation flux  $F$ 
(in units 10$^{10}$ phot cm$^{-2}$ s$^{-1}$ eV$^{-1}$ at
the Lyman limit) which is defined for AGNs  and the starburst parameters (the ionization parameter $U$ and 
the temperature of the stars (in units of 10$^4$ K));
bottom left : the relative abundances in units of 10$^{-4}$; 
bottom right : \Haa and $D$.
}
\end{figure*}

\begin{figure*}
\includegraphics[width= 5.8cm]{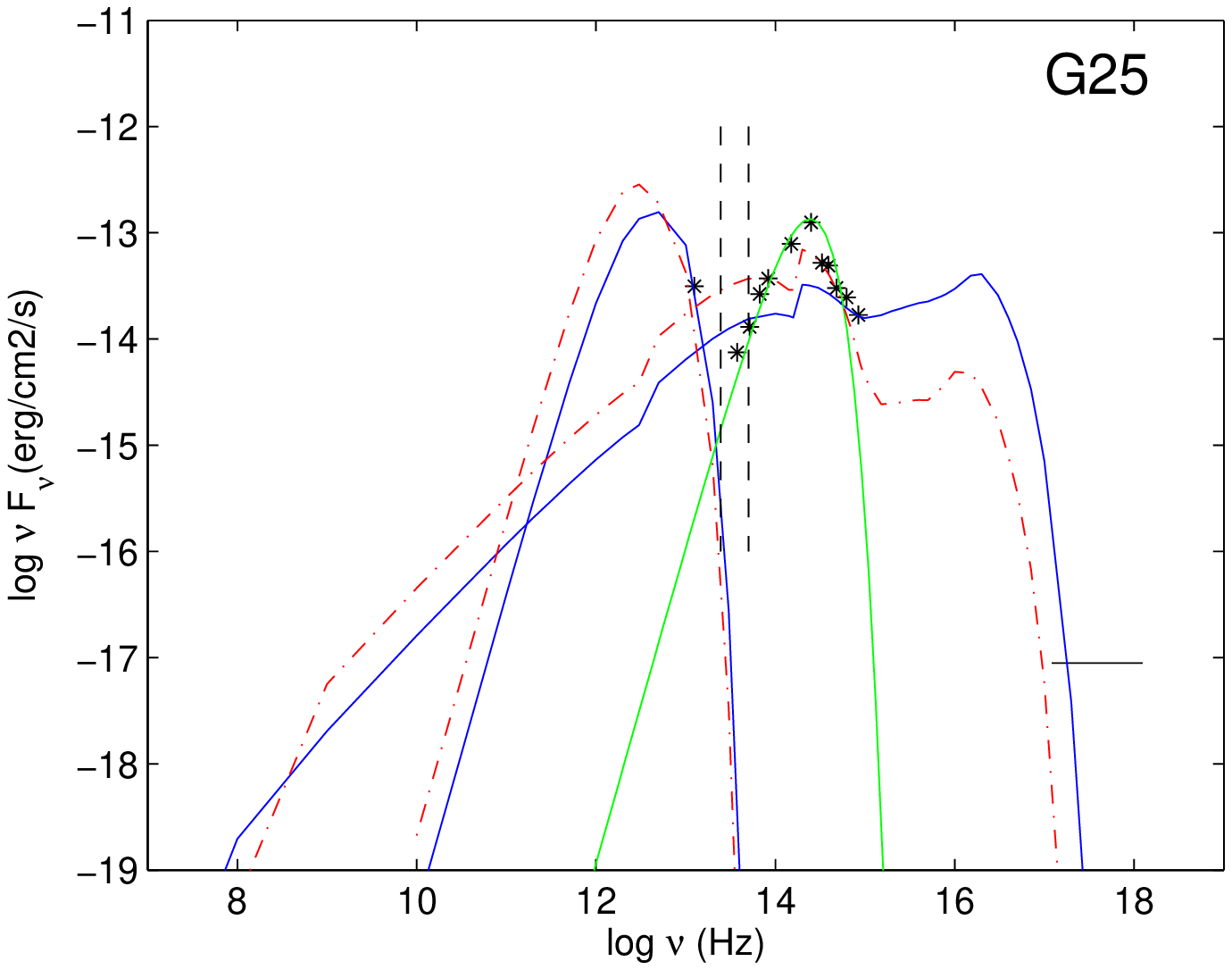}
\includegraphics[width= 5.8cm]{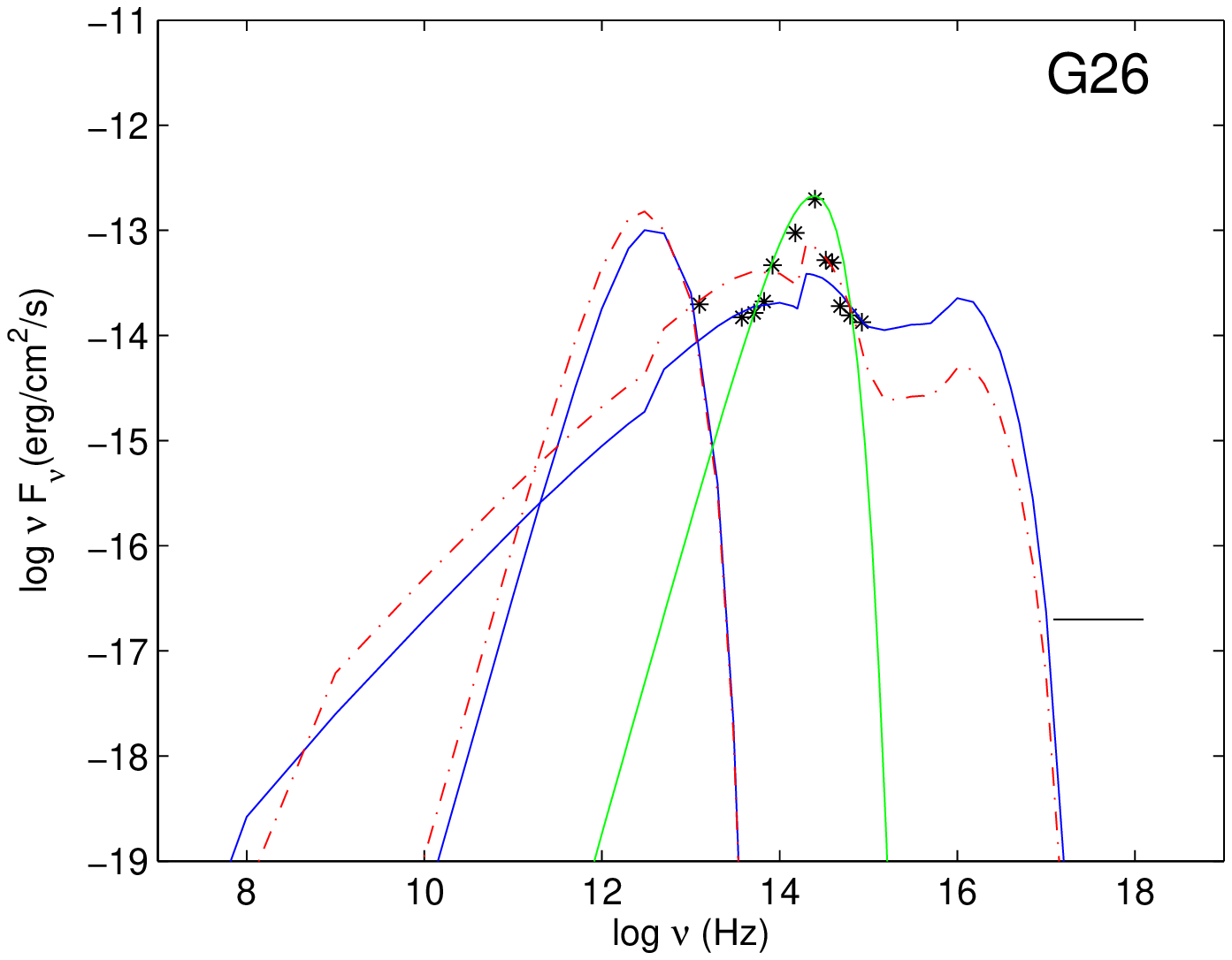}
\includegraphics[width= 5.8cm]{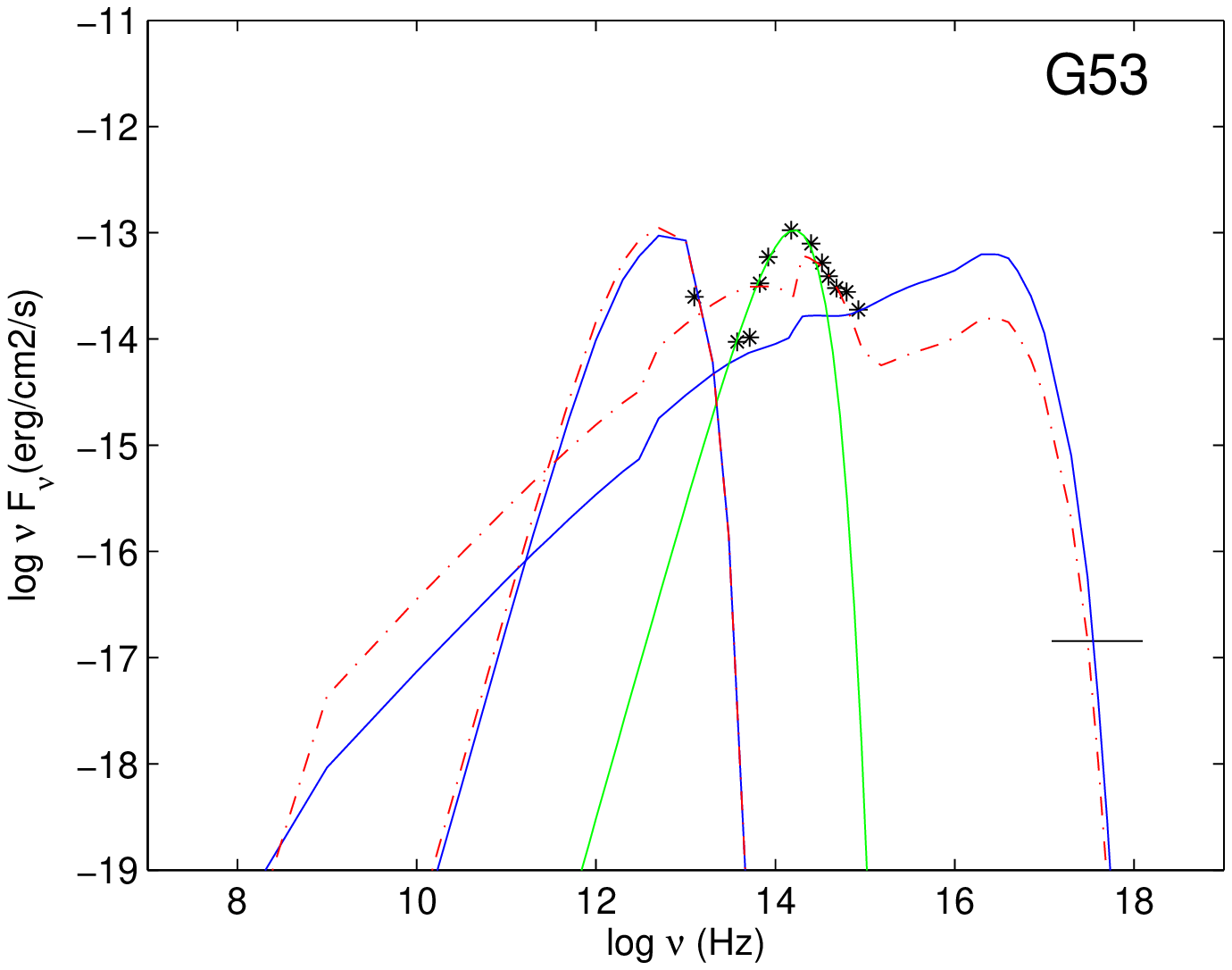}
\includegraphics[width= 5.8cm]{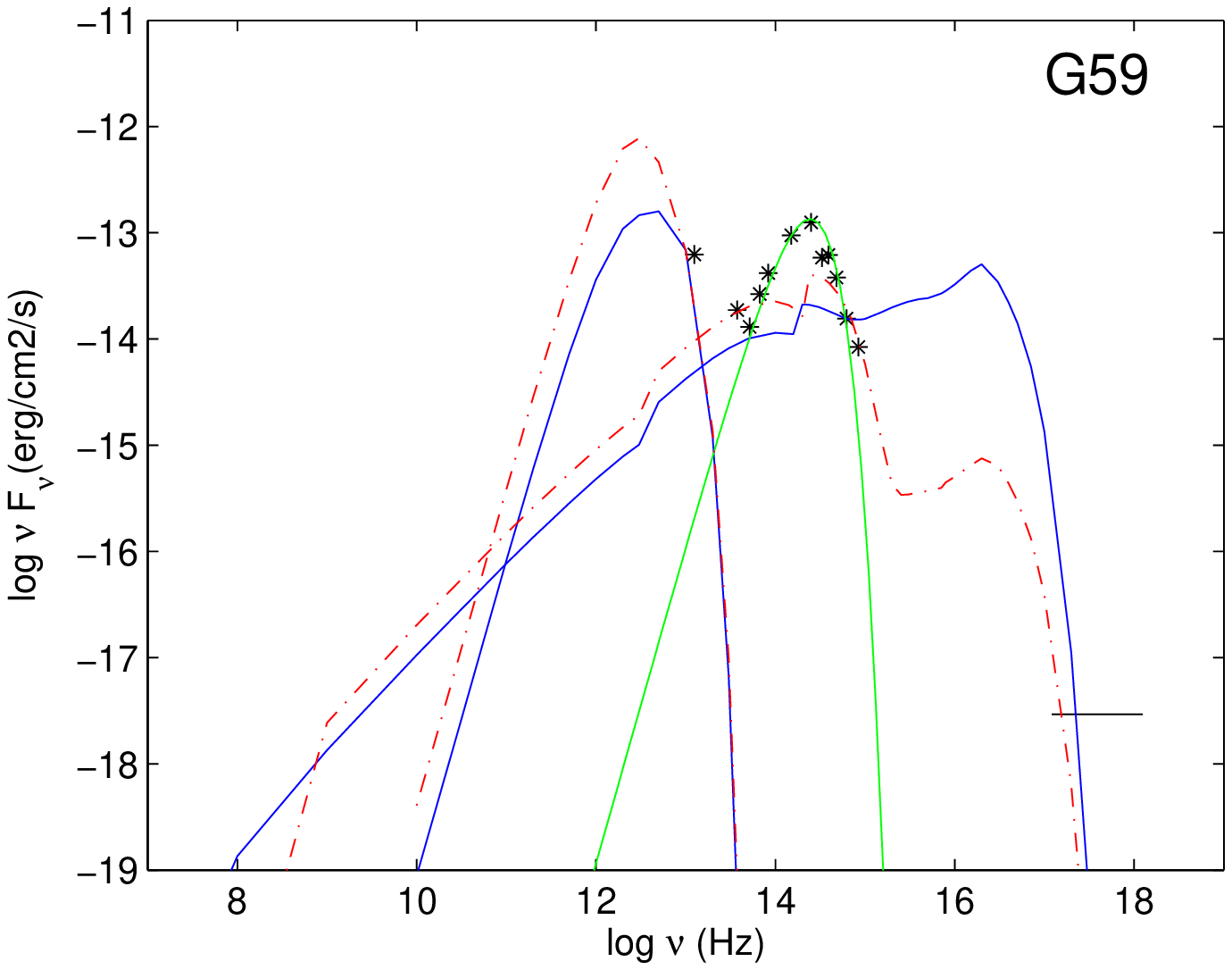}
\includegraphics[width= 5.8cm]{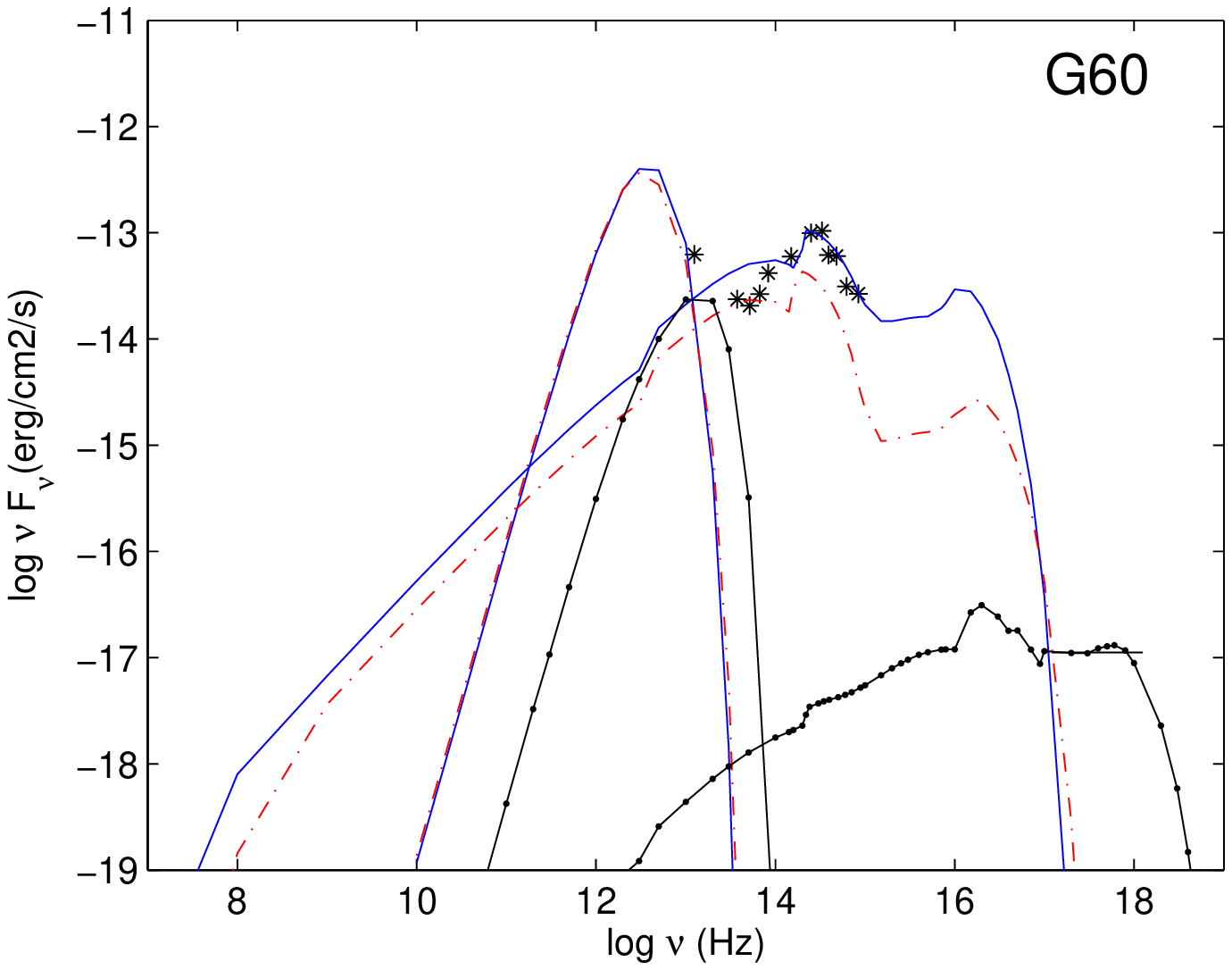}
\includegraphics[width= 5.8cm]{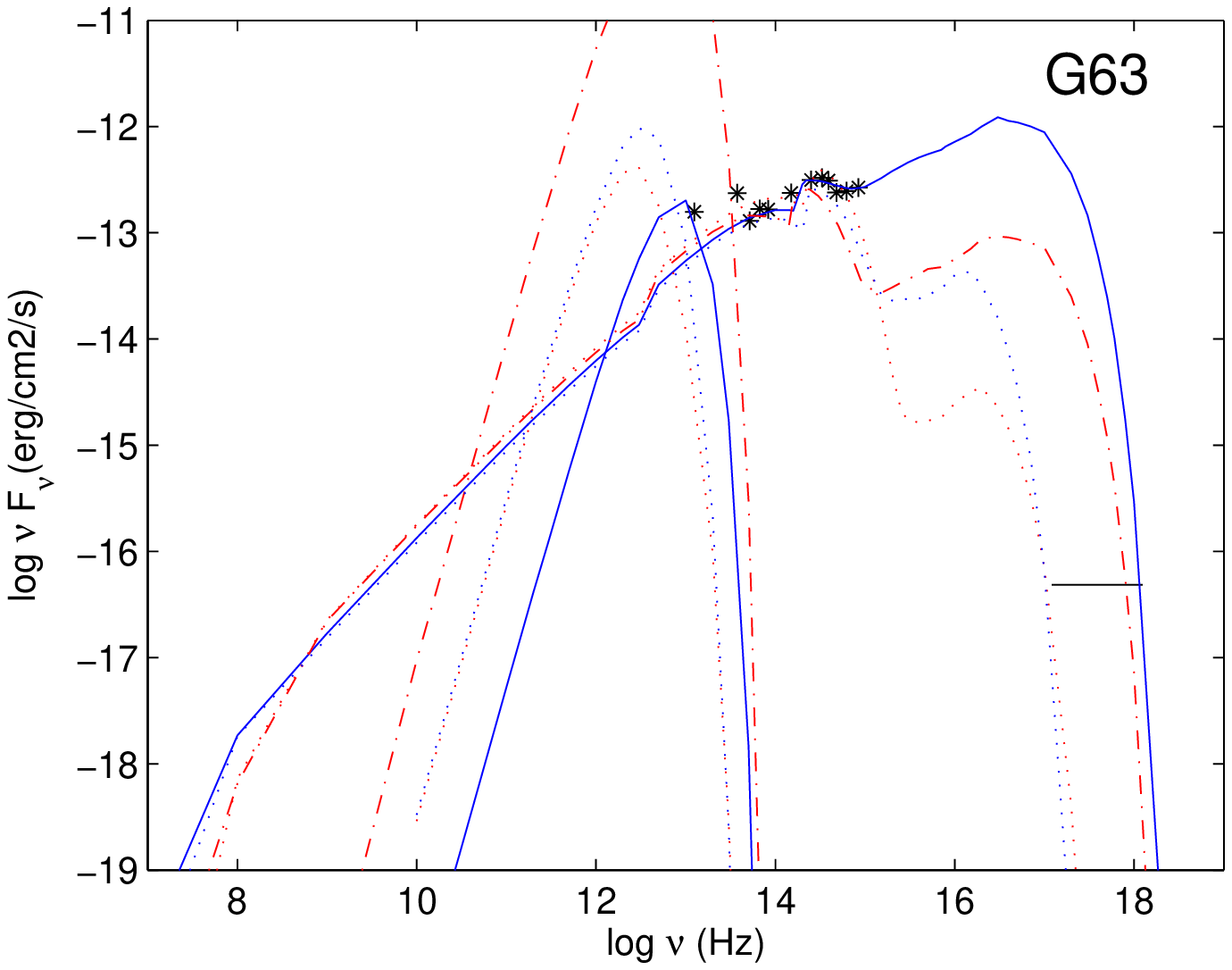}
\includegraphics[width= 5.8cm]{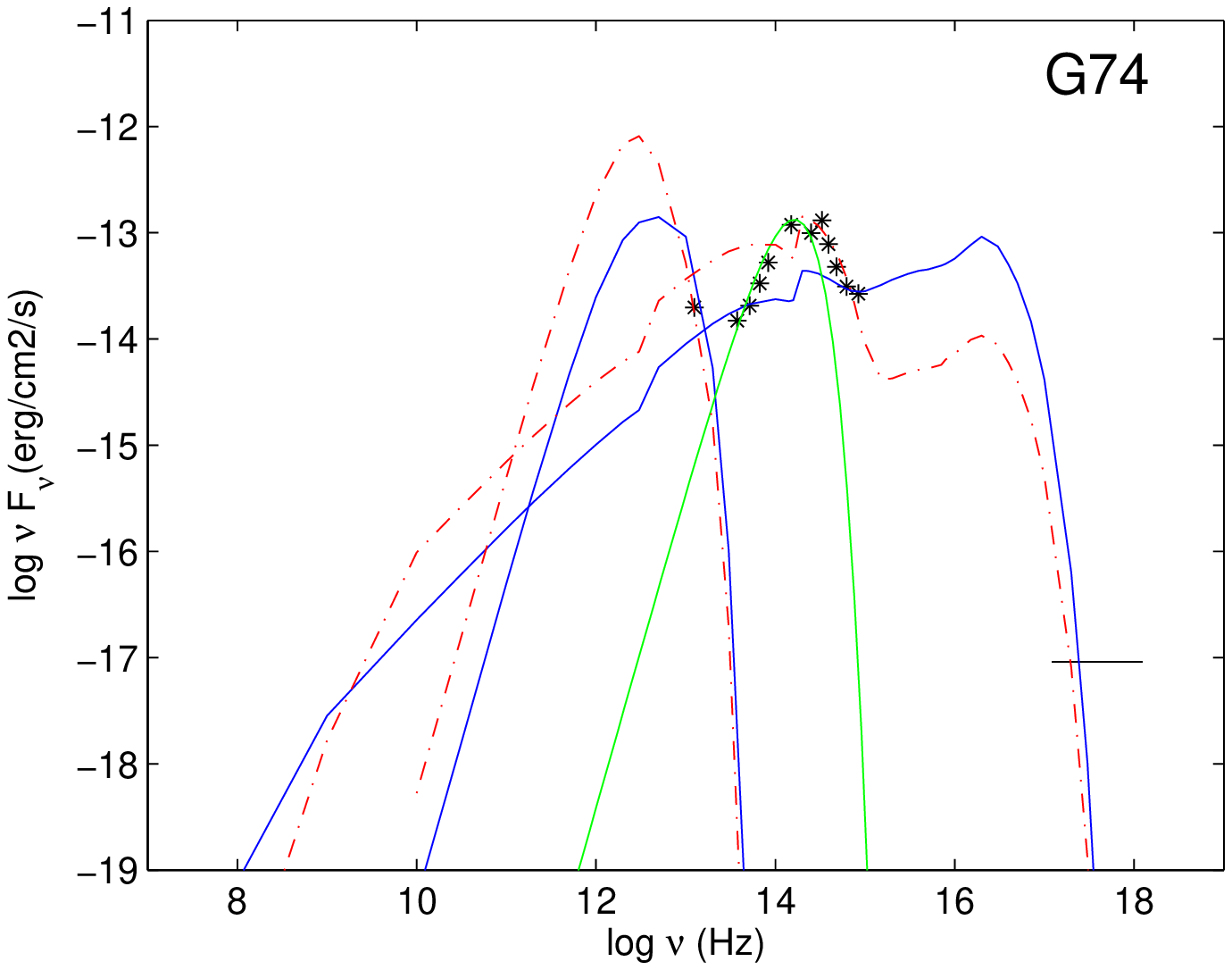}
\includegraphics[width= 5.8cm]{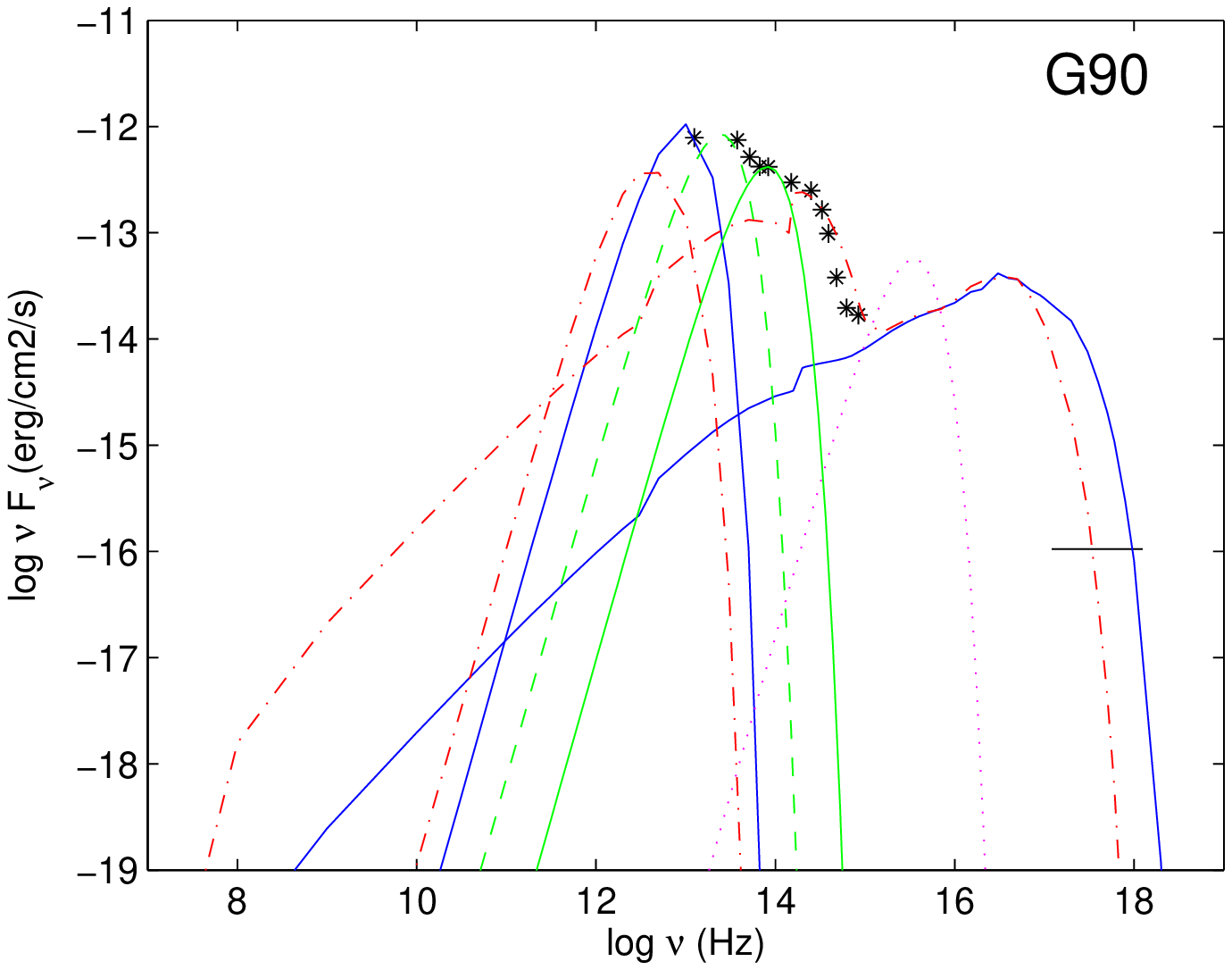}
\includegraphics[width= 5.8cm]{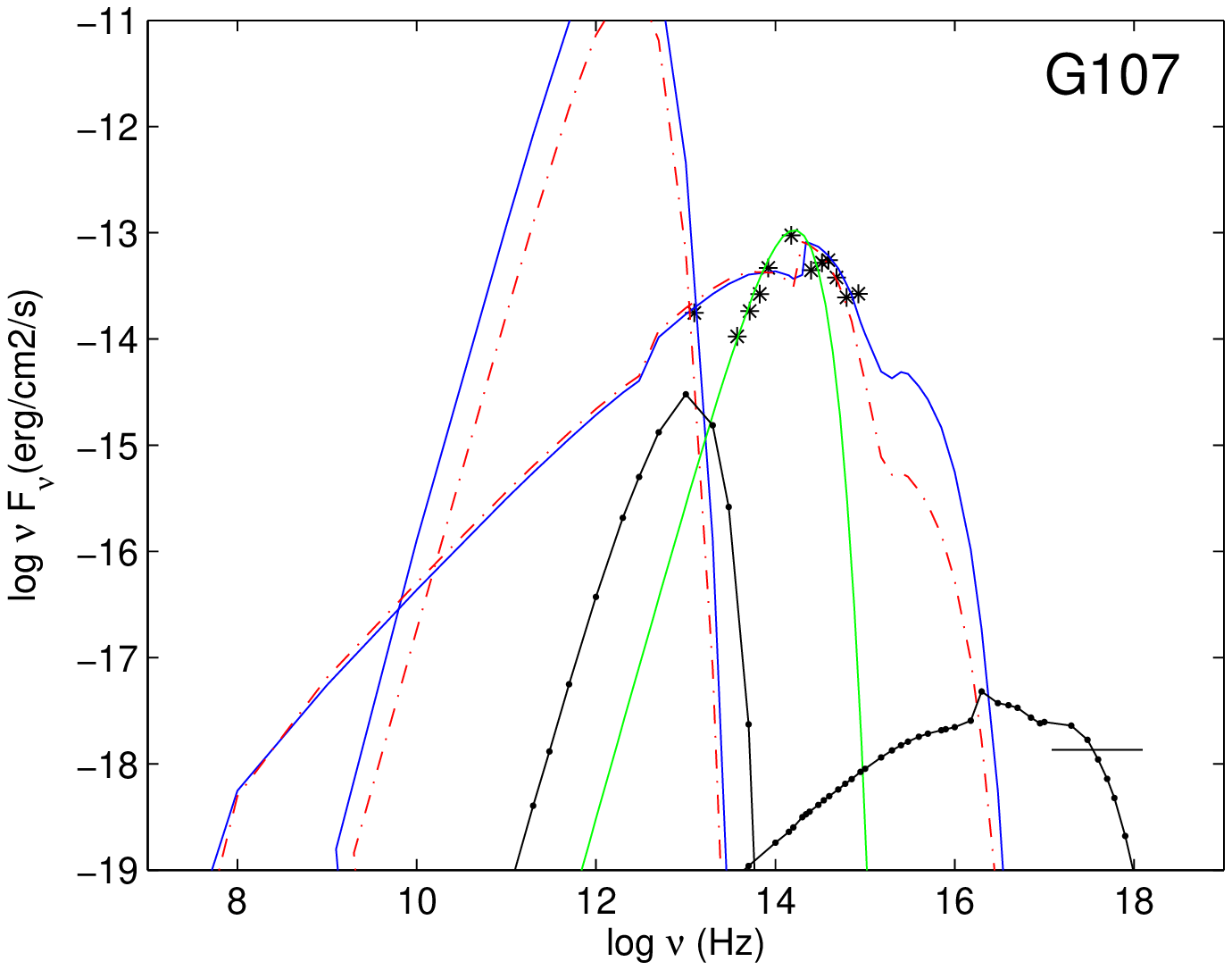}
\caption{The best fit of model calculations to the RA13  continuum data.
Black asterisks : the data; black horizontal segment : the X-ray flux; blue line :  AGN model
for G63b; 
red dot-dashed line : starburst model for G63b; blue dotted : AGN model for G63r;
red dotted lines : SB model for G63r
black  lines marked with dots : shock dominated high velocity models ;
green lines : black body radiation ; magenta  dotted line : the black body flux from the starburst stars.
}

\end{figure*}

\begin{table*}
\caption{The  parameters used to model the continuum SED}
\begin{tabular}{cccccccccc} \hline  \hline
\    & d$^1$   & $\eta_{AGN}$& r$_{AGN}$$^2$ & $\eta_{SB}$ & r$_{SB}$$^2$ & T$_{bb}$   & type$^3$\\ \hline
\ G25&3.04  &-12.       & 3.         &-12.5 & 1.7        & 3000      & NG  \\
\ G26&3.23  & -11.9    & 3.6        &-12.5 & 1.8       & 3000    & SB dom \\
\ G53&2.88  & -12.1    & 2.6        &-12.7 & 1.29       & 2000     & Type 2 \\
\ G59&1.86  & -11.6    & 2.95       &-13.5 & 0.33       & 3000     & NG      \\
\ G60&1.94  & -11.4    & 3.87       & -12.9& 0.69       &  -       & Type 2 \\
\ G63b&1.93  & -11.1    & 5.44       & -12.1& 1.7       &  -       & Type 1  \\
\ G63r&1.93  & -10.9    & 6.8           & -12.5&1.08    &  -        & Type 1  \\
\ G74&2.20  & -11.9    & 2.47       & -12.6& 1.1       & 2000     & NG  \\
\ G90&4.59  & -13.     & 1.45       & -11.9& 5.1       & 1000     & SB cont     \\
\   &         &          &            &      &           & 300         &      \\
\ G107&2.68 & -11.2    & 6.73       & -11.95& 2.84      & 2000        & NG    \\ \hline
\end{tabular}

$^1$ in 10$^3$ Mpc ;
$^2$ in kpc ;
$^3$ NG (normal galaxy); SB dom (SB dominated); Type 2 (Seyfert 2 like); Type 1 (Seyfert 1 like);
SB cont (SB contaminated) (RA13, sect. 2).

\end{table*}

\subsection{Results obtained by   modelling  the RA13 line spectra}

We consider that both the AGN and the  SB coexist in the RA13 sample galaxies.
The parameter sets which best reproduce the observed line ratios for each galaxy
are presented  in Tables 3, 4 and Fig. 4.
  The tables show that the preshock densities are rather high ($\sim$ 1000 \kms) 
 compared to those of  galaxies at  z $\leq$0.1 .
In particular, they resemble  \n0 
  calculated in the NGC 3393 merger Seyfert 2 galaxy at z=0.0125. It was explained by Contini (2012a)
that relatively high preshock densities may indicate that a  shock wave   
 originating  at collision of  merging galaxies has compressed the gas throughout the  NLR.

The preshock densities in the present galaxy sample follow an increasing trend with increasing z.
The range of the shock velocity  is the same
for AGN and SB  dominated models because  the velocities  were calculated  on the basis of 
the FWHM of the line profiles.

The photoionizing flux from the AGN is rather  constant but the ionization parameter calculated by the starburst models
 decreases  for z increasing from  0.4 to  0.8.

The O/H relative abundances are mostly solar  (adopting 6.6 10$^{-4}$ Allen 1976,  8.5 10$^{-4}$ refers to
Anders \& Grevesse 1989) 
in  the gas
heated and ionized by the starburst (Table 4), while
 N/H ranges from 3 times solar
(G63 and G90) to a minimum of 0.2 solar in G107.
Some of the galaxies (G25, G26) show
O/H  $<$ 0.5 solar   in the neighborhood of  AGNs
and almost solar  O/H in other AGN except for G53 and G107 at z=0.72 and 0.67, respectively,
which also show 
N/H lower than solar by  factors of 3 and 5.  
The gaseous clouds  present in average a higher fragmentation (indicated by a smaller geometrical thickness $D$) 
in the NLR of the AGN than in the regions close to the  starburst. For both AGN and SB, $D$ decreases for z  increasing from 0.4 to 0.8.
The \Ha~ absolute fluxes calculated at the nebula are lower by a factor of $\leq$ 100 for AGNs than for the SBs. 

 G62   shows the characteristic parameters  different from the bulge observed from the other sample objects (Fig. 3),
but similar to those  observed by Kobulnik \& Zaritsky (1999) from HII regions in
compact galaxies (see Contini 2013c, in preparation), so it  has been modelled only by a starburst dominated model.

We are dealing with a sample of only ten  galaxies. 
We  will  compare   the results obtained for the RA13 sample at intermediate redshifts with those
 obtained for galaxies at lower z,  such as
  local merger  galaxies, LINERs and local starburst, etc. as well as for galaxies  at higher z in an accompanying paper 
(Contini 2013c, in preparation).

\subsection{The continuum SEDs}

In the previous sections we have constrained the  characteristic parameters of the RA13 sample galaxies by
modelling the line spectra. 
We have referred to models  for both
 AGN  and starburst  galaxies. We can now  analyse the continuum SEDs by the same models.
The fit of the data throughout the continuum SED 
 by both the AGN and SB  models, can give a hint about the relative importance of the AGN and the SB 
in each of the objects. 

The  templates  adopted by RA13 in their fig. 2  to explain the observed IR data, were 
obtained combining the observed  continuum SEDs of
many objects  for  each of the  sample galaxy.
RA13 classified the galaxies as 
starburst dominated, starburst contaminated, Seyfert 1, Seyfert 2 and  normal galaxies 

In the present investigation we  use a different method, namely, we  model  each galaxy continuum SED  consistently 
with the models that were constrained reproducing the line ratios.
We present our results in Fig. 5, where we show  the continuum SEDs on a large frequency range  ($\nu$ 
between 10$^7$ and 10$^{19}$ Hz) 
 including  the radio and X-ray ranges. The  radio data are not given by the observations but they can be 
predicted by the bremsstrahlung calculated by the models. On the other hand,  if the radio flux
is synchrotron radiation created by the Fermi mechanism at the shock front,  a different slope would be observed.

There is now  common agreement that there is  mutual triggering between the starburst  and  the AGN
activity in the galaxies. So in  Fig. 5 diagrams we have included both kinds of models.
The data are observed at Earth, while the models are calculated at the nebula. Therefore, model results
are compared with the data adopting a readjusting factor $\eta$ (r$^2$=$\eta$ d$^2$ \cf) which accounts for 
the distance of the
emitting nebula from the galaxy centre (r),  for the distance of the galaxy from Earth (d) and for the 
cover factor \cf. They are reported in Table 5. 

 In the Fig. 5 diagrams two curves appear for each of the models.
One  represents the bremsstrahlung emitted from the nebula and the other, peaking in the IR,
represents the reprocessed radiation from dust, calculated consistently with the gas continuum emission.
The dust grains are heated  by collision with the gas and by radiation from the active source.
At high temperatures collisional processes dominate, therefore the higher the shock velocity,
the higher the frequency of the dust radiation  peak.
Fig. 5 shows that the datum at 24 \mum is always reproduced by dust reprocessed radiation.  
Presently it is the only datum 
for $\lambda$ $>$ 10 \mum,
so the peak of the dust reprocessed radiation flux is constrained by  one only  datum. 
Dust-to-gas ratios d/g= 4 10$^{-3}$ were adopted in the present models.
The $\eta$ factors for the dust reradiation peak are lower than for those of the bremsstrahlung 
by factors $<$ 10. We suggest that the dust has a patchy distribution. 
At frequencies referring to wavelengths between $\sim$ 8 and 12 \mum the large crater of silicate absorption
at $\lambda$ $\sim$ 10 \mum
should be taken into consideration. Moreover, there is large absorption by ices and HAC at 6 \mum and 7 mum
respectively (Spoon et al 2002). They are included into the band defined by the vertical dashed lines in Fig. 5
(top left panel). 

For all the sample galaxies, except for G60, G63 and G90, a black body radiation curve calculated by a single 
temperature T$_{bb}$
contribute to the modelling of nearly all the observed data by RA13. The temperature ranges between 2000 and 3000 K
for all the galaxies except for G90  where T$_{bb}$ = 1000 K and T$_{bb}$ = 300 K  are required to  finish the
fit of the NIR and FIR data,
respectively (Table 5). We suggest that a temperature of 1000 K represents the low temperature of the background stars,
while a temperature of 300 K can be achieved by dust characterized by large grains which could survive sputtering by 
a high velocity shock (\Vs$\geq$ 1000-2000 \kms). 

The bremsstrahlung from the  high \Vs gas contributes to the
radiation maximum in the X-ray domain.
We have added in Fig. 5  the X-ray fluxes calculated from RA13. 
It can be seen that in  G53, G59, G63, G74 and G90 the models  calculated by  \Vs $>$ 200 \kms nicely fit the data,
G25 and G26 are at the limit. The two galaxies show components with large FWHM that  translate  to
velocities $>$ 1500 \kms and $>$ 800\kms, respectively.  Those components refer  only to \Ha~ and \Hb,
indicating  some contribution from the broad line region.
Therefore the spectra cannot be properly reproduced with the present models. 

Notice that the  high frequency region in the SEDs
is determined by the shock velocity. The  stronger the shock, the higher the temperature in the immediate
post shock region downstream. Therefore we added in G60 and G107 diagrams the SED of shock dominated
models calculated  with a high \Vs in order to fit the X-ray data. In G26 the high shock component is not observed.
Perhaps the broad profile  of weak lines is hidden by the continuum noise.

In Table 5 we summarize the results which appear  from Fig. 5. In column 2 the distances of the galaxies in Mpc are given,
The $\eta$ factors followed by the radius r
calculated for AGNs  and SBs, respectively, are presented in columns 3-6. In column 7 the temperature of the
background stars is  reported followed by the galaxy  classification type by RA13.

\section{Discussion and concluding remarks}

\begin{figure}
\includegraphics[width=8.6cm]{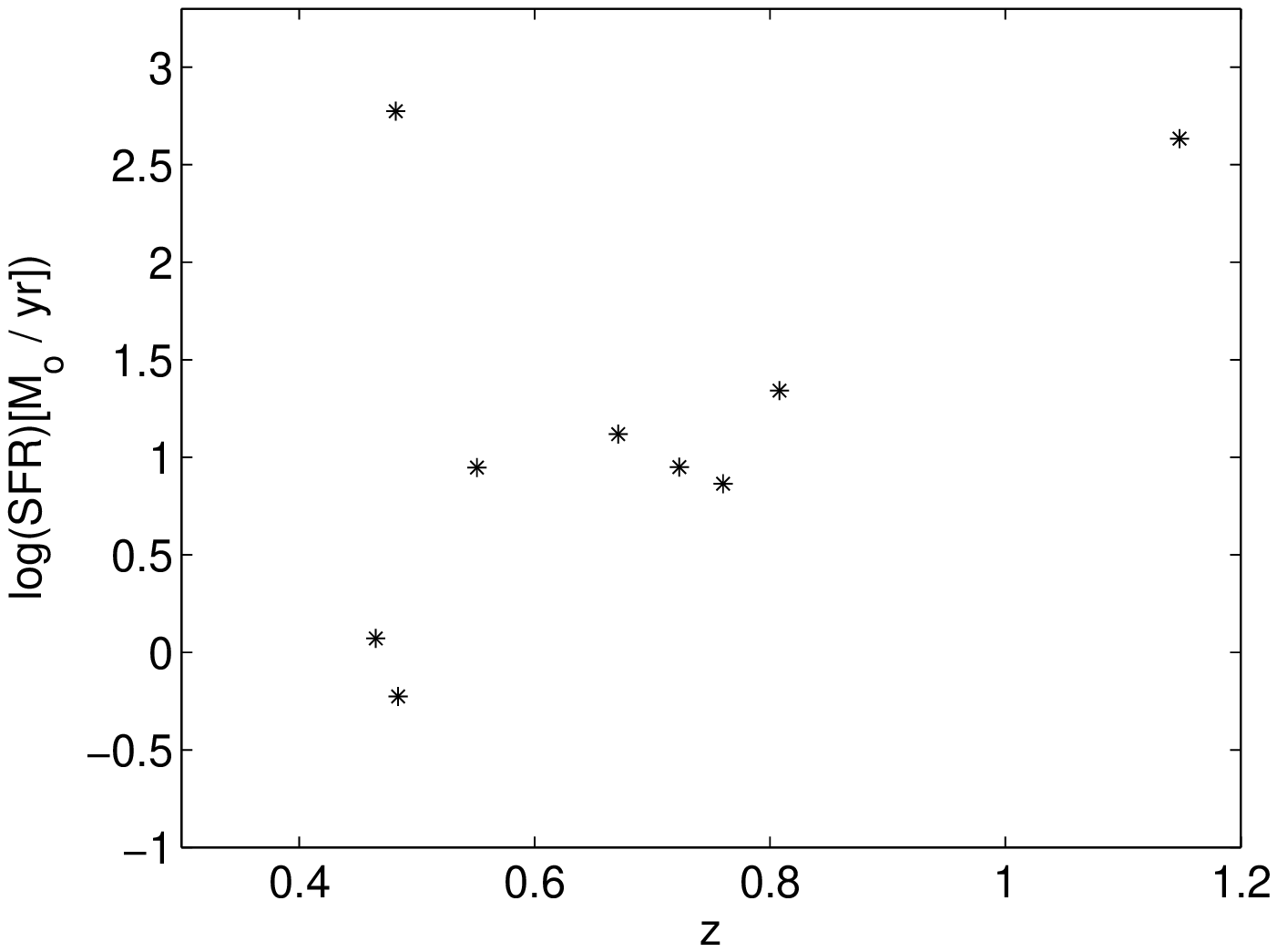}
\caption{SFR as a function of z from RA13, table6}
\includegraphics[width=8.6cm]{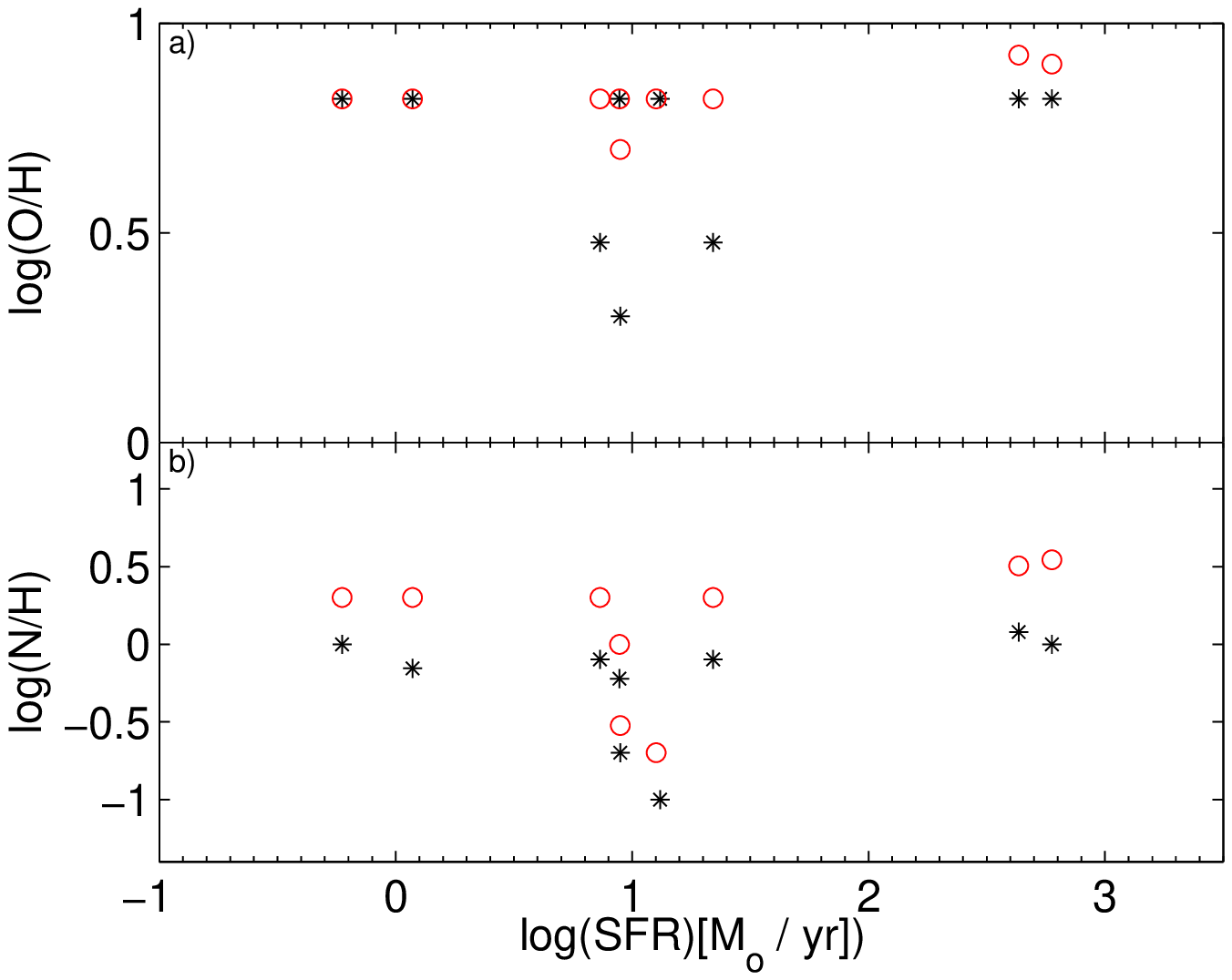}
\caption{Metallicity as a function of SFR}
\end{figure}

We have analysed the spectra presented by RA13 for a sample of galaxies at redshift 0.27$<$z$<$ 1.28.
Only  ten  galaxies were selected because they showed line ratios which could constrain
the modelling interpretation.

The results of model calculations show that G59, G60, G63 are most likely  AGN dominated, G62 is SB dominated
while the other galaxies are ambiguously defined by the line ratios presented in RA13 spectra.
The continuum SEDs show that most of the data cover the frequency range corresponding to the old star background
with temperature of 2000-3000 K therefore the modelling is less defined. 
The SEDs confirm that G60 and G63 are AGN dominated. G26 seems also AGN dominated.
Most of the objects in the RA13 sample show that an AGN and a SB coexist in the galaxy.

We were particularly interested in the distribution of the emitting gas physical conditions and relative abundances
throughout the redshift.
Fig. 4 shows that the shock velocities and the preshock densities have opposite trend in agreement with Rankine-Hugoniot
equation for mass conservation at the shockfront. A dip  at z between 0.6 and 0.8 appears 
for \Vs, $U$, $D$ and for
the N/H and O/H relative abundances indicating that  an interfering process has disturbed the evolution of the
parameter trends. Notice for instance that  fragmentation of the clouds in terms of $D^{-1}$ close to the starburst is  gradually decreasing
with z decreasing from z= 0.8 to  z=0.6. 
In the meanwhile the N/H abundance dropped, while O/H decreased but less drastically.

 RA13  presented the results  of the  analysis of the spectra  in their table 6.
In Fig. 6 we report RA13 calculations  showing the SFRs as  function of the redshift.
 In the redshift range covered by the sample selected  by our models, SFRs are  increasing with z, except for G63b. 
Notice that the errors
which appear in RA13 table 6 are very large.

Finally we  show in Fig. 7
the metallicities as a function of SFR. There are a few objects
 which show relatively low O/H relative 
abundances at SFR  $\sim$  10 \msol  yr$^{-1}$.
Also N/H has a dip in this SFR range.
This could suggest that a process such as  strong star formation is starting.

The results of our modelling require a deeper  interpretation of the data, which will be possible comparing
the results obtained for  this galaxy sample  with  the results which will be obtained by modelling different galaxies 
throughout a larger frequency range.

% \begin{thebibliography}{99}
\section*{References}

\def\ref{\par\noindent\hangindent 18pt}
\ref Allen, C.W. 1976 Astrophysical Quantities, London: Athlone (3rd edition)
\ref Anders E., Grevesse N. 1989, Geochim. Cosmochim. Acta, 53, 197
\ref Baldwin J. A., Phillips M. M., Terlevich R.  PASP, 1981, 93, 5
%\ref Cardelli, J.A., Clayton, G.C., Mathis, J.S.  1989, ApJ, 345, 245
\ref Collins, N.R., Kraemer, S.B., Crenshaw, D.M., Bruhweiler, F.C., Mel\'{e}ndez, M. 2009, ApJ, 694, 765
\ref Contini, M. 2013a, MNRAS, 429, 242
\ref Contini, M. 2013b, MNRAS, submitted
\ref Contini, M. 2012a, MNRAS, 426, 719
\ref Contini, M. 2012b, MNRAS, 425,120
\ref Contini, M. 2009, MNRAS, 399, 1175
\ref Contini, M. 1997, A\&A, 323, 71
\ref Contini, M., Aldrovandi, S.M. 1983, A\&A, 127, 15
\ref Contini, M., Cracco, V., Ciroi, S., La Mura, G. 2012, A\&A, 545, 72 
\ref Contini, M., Viegas, S.M. 2001, ApJS, 132, 212
%\ref Contini, T., Treyer, M.A., Sullivan, M., Ellis, R.S. 2002, MNRAS, 330, 75
\ref Gunawardhana, M.L.P. et al. 2013MNRAS.tmp.1654
\ref Izotov, Y. I.; Stasińska, G.; Meynet, G.; Guseva, N. G.; Thuan, T. X.  2006, A\&A, 448, 955
\ref Izotov, Y. I.; Guseva, N. G.; Fricke, K. J.; Stasińska, G.; Henkel, C.; Papaderos, P.
2010, A\&A, 517, 90
\ref Kauffmann, G. et al. 2003, MNRAS, 346, 1055
\ref Kewley, L.J., Dopita, M.A., Sutherland, R.S., Heisler, C.A., Trevena, J. 2001, ApJ, 556, 121 
\ref Kobulnicky, H., Zaritsky, D. 1999, ApJ, 511, 118
\ref Osterbrock, D.E. 1974  in "Astrophysics of Gaseous Nebulae" .  ed. W.H. Freeman and Company , San Francisco
\ref Ramos Almeida, C., Rodr\'{i}guez Espinosa, J.M., Acosta-Pulido, J.A. Alonso-Herrero, A., 
P\'{e}rez Garc\'{i}a, A.M, Rodr\'{i}guez-Eugenio, N. 2013, MNRAS, 429, 3449
\ref Spaans, M. \& Carollo, C.M. 1997 ApJ, 482, L93 
\ref Spoon, H. W. W.; Keane, J. V.; Tielens, A. G. G. M.; Lutz, D.; Moorwood, A. F. M.; Laurent, O. 2000, A\&A, 385, 1022
\ref Springel, V., Di Matteo T., Hernsquist, L. 2005, ApJ, 620, L79

\end{document}